\documentclass[10pt,a4paper,useAMS,usenatbib]{article}
\usepackage{jcappub}
\usepackage[english]{babel}

\usepackage{fancyhdr}
\usepackage{amsfonts}
\usepackage{amsmath}
\usepackage{amssymb}
\usepackage{multicol}
\usepackage{layout}
\usepackage{graphicx}
\usepackage{epstopdf}

\usepackage{times}
\usepackage{natbib}

\newif\ifAMStwofonts
\AMStwofontstrue

\title{The clustering of baryonic matter. I: a halo-model approach}

\author[a,b]{C. Fedeli}
\affiliation[a]{INAF - Osservatorio Astronomico di Bologna, via Ranzani 1, 40127 Bologna, Italy}
\affiliation[b]{Department of Astronomy, University of Florida, 211 Bryant Space Science Center, Gainesville, FL 32611}
\emailAdd{cosimo.fedeli@oabo.inaf.it}

\abstract{In this paper I generalize the halo model for the clustering of dark matter in order to produce the power spectra of the two main baryonic matter components in the Universe: stars and hot gas. As a natural extension, this can be also used to describe the clustering of all mass. According to the design of the halo model, the large-scale power spectra of the various matter components are physically connected with the distribution of each component within bound structures and thus, ultimately, with the complete set of physical processes that drive the formation of galaxies and galaxy clusters. Besides being practical for cosmological and parametric studies, the semi-analytic model presented here has also other advantages. Most importantly, it allows one to understand on physical ground what is the relative contribution of each matter component to the total clustering of mass as a function of scale, and thus it opens an interesting new window to infer the distribution of baryons through high precision cosmic shear measurements. This is particularly relevant for future wide-field photometric surveys such as \emph{Euclid}. In this work the concept of the model and its uncertainties are illustrated in detail, while in a companion paper we use a set of numerical hydrodynamic simulations to show a practical application and to investigate where the model itself needs to be improved.}

\begin{document}
\maketitle

\section{Introduction}\label{sct:introduction}

It is nowadays widely accepted that $\sim 80\%$ of all matter in the Universe is non-baryonic in nature, and composed by a cold thermal relic \cite{RU80.1,TE04.1,BE05.1,DI11.1,KO11.1} dubbed Dark Matter (DM henceforth). Several lines of investigation strongly suggest that DM particles have a very low scattering cross section \cite{MA04.2,RA08.1,AN13.1,ZA13.1}, meaning that, at an excellent level of approximation, DM interacts only gravitationally. That being the case, theoretical modeling of the growth of DM structures is a demanding, yet relatively straightforward undertaking. Gravity is a scale-free interaction, which makes it particularly suitable for numerical modeling. Over the past two decades a substantial amount of effort has been put into simulating the clustering of DM in the Universe by means of $n-$body simulations \cite{DU91.1,NA95.1,NA96.1,NA97.1,SM03.1}, and a standard theoretical scenario has emerged which compares well with cosmological observations. In spite of the many details still in need to be worked out, there is now widespread consensus on how DM should be distributed in space.

The situation is radically different for the remaining $\sim 20\%$ baryonic matter component, which is mainly made up by gas and stars. Despite its large-scale dynamics being dominated by gravity, gas is capable of exchanging energy with radiation backgrounds \cite{KA91.1}, thus altering its own thermodynamics. Moreover, on small scales gas has a finite pressure, which leads to radiative cooling. Stars, that fragment out of cold gas clouds, can be treated as collisionless matter of the like of DM. However, at the end of their life cycle stars inject both energy and metals back into the gas, thus altering its entropy content and radiative properties. Energy feedback from Active Galactic Nuclei (AGN), fueled by gas accretion on their central supermassive black holes, is also expected to play a major role in the shaping of baryons in the Universe \cite{MC10.1,MC11.1,VA11.1}. The interplay of all these non-gravitational physical processes is very resilient to theoretical modeling (see e.g., \cite{WU13.1} for a recent review on the difficulty of simulating AGN activity), and a comprehensive understanding of the spatial distribution of gas and stars is still far from being obtained. On top of that, non-gravitational physical processes are important on such small scales that are impossible to be resolved in a cosmological simulation. The end result is that literature on the clustering of baryons offers varied and sometimes contradictory conclusions (see the reviews of past works on this topics in \cite{FE11.1} and \cite{VA11.1}).

Crucial information about the clustering of matter down to small separations is contained in the non-linear power spectrum, which provides a measure of the variance of density fluctuations as a function of scale. Measurement of the matter power spectrum has emerged as a tool to infer essential clues about cosmology and the assembly of cosmic structure, and it is thus a key objective of current and future cosmological investigation. One particularly effective way to perform these measurements is via cosmic shear \cite{BA00.1,BA01.1,HO08.2,MU08.1}, i.e. the correlation of galaxy image distortions produced by gravitational lensing across the celestial sphere. The gravitational deflection of light is unable to distinguish different matter components, thus for a given source redshift distribution cosmic shear produces a projected map of the \emph{total} mass distribution in the sky. Class III cosmic shear experiments (see the nomenclature adopted by the Dark Energy Task Force \cite{AL06.1}), such as the Dark Energy Survey \cite{TH05.1} are under way, while Class IV experiments are being developed, which will allow a measurement of the total matter power spectrum with increasingly high precision. The ESA Cosmic Vision mission \emph{Euclid}\footnote{http://www.euclid-ec.org/} \cite{LA11.2} is an example of a Class IV experiment. Scheduled for launch in $2020$, it is expected to measure cosmic shear over $15,000$ square degrees, quantifying the matter clustering with a precision of a few percent.

Despite the wide uncertainties plaguing the results of cosmological simulations, it is rather well established that the impact of non-gravitational physics on the matter power spectrum is much larger than a few percent. Different authors find modifications of the order of $\sim 5-50\%$ at a wavenumber $k \sim 5 h$ Mpc$^{-1}$ \cite{JI06.1,RU08.2,GU10.1,CA11.1,FE11.1,VA11.1}, and up to a factor of $\sim 2$ at $k \sim 50 h$ Mpc$^{-1}$. This effect is not only due to the fact that gas and stars behave differently from DM, but also to the fact that baryons and DM are gravitationally coupled. This means that non-gravitational physics cause a rearrangement of the DM distribution (\emph{backreaction}) as well. It is thus clear that our imperfect theoretical knowledge of the matter clustering constitutes a substantial systematic uncertainty that hinders the full exploitation of future cosmic shear experiments \cite{AN05.1}. In order to mitigate this problem it is important that we gain a more coherent understanding of the spatial distribution of \emph{all} matter in the Universe, or else that we find a meaningful procedure to parametrize such an uncertainty so that it can be later marginalized over. The latter approach would be equivalent at reducing the impact of systematic uncertainties on cosmological parameter estimation at the expense of larger statistical errors.

This kind of investigation has been undertaken recently in a simplified way by several authors \cite{SE11.1,ZE13.1}. In this paper I discuss a more complete and internally consistent procedure, lying in the development of a Semi-Analytic Model (SAM) that describes, individually, the clustering of each of the three main matter components in the Universe: DM, hot gas, and stars. In this context, the power spectrum of each matter component can be related to its abundance and distribution within the gravitational potential wells of DM halos, allowing the SAM to have a direct and an inverse application. The direct application consists in feeding the model with the observed baryon and DM distributions within galaxies and galaxy clusters and then analyze the resulting total matter power spectrum. This application allows one to study the effect of baryons on cosmological parameter estimation, a task that cannot be accomplished with hydrodynamic simulations, and to parametrize the relative contribution of each matter component to the total mass clustering as a function of scale. Moreover, in this way one can see how cosmological forecasts change as one either marginalizes over such contributions, or updates the baryon distribution as observations of the internal structure of cosmic objects become increasingly more accurate. The inverse application consists in using the total matter power spectrum (e.g., measured through cosmic shear), in order to gain insights on the mass fractions and density profiles of the various matter components within DM halos. Since these distributions depend ultimately on the physical processes responsible for the formation of galaxies and galaxy clusters, one could effectively use cosmic shear observations to learn about these physical processes. Moreover, interpreting cosmic shear observations in terms of the SAM would permit to place constraints on the distribution of matter components that are very difficult to gauge otherwise, for instance intra-halo stars and the diffuse gas on super-halo scales.

In this paper I present the general layout of the SAM. A companion paper (Fedeli et al. in preparation, Paper II henceforth) shows an application of the model based on numerical hydrodynamic simulations implementing different kinds of baryonic physics, in order to investigate the accuracy of the SAM and how it can be improved. In particular, we shall make use of the OverWhelmingly Large Simulation (OWLS) project \cite{SC10.1}. For consistency, in this work I adopted the cosmological parameter set of the OWLS project, however the SAM is by no means restricted to such a set. Accordingly, I used $\Omega_{\mathrm{m},0}=0.238$ for the total matter density, $\Omega_{\Lambda,0}=1-\Omega_{\mathrm{m},0}$ for the vacuum energy density, $\Omega_{\mathrm{b},0}=0.042$ for the baryon density, $h\equiv H_0/(100 h$ km s$^{-1}$ Mpc$^{-1}) = 0.730$ for the Hubble constant, $\sigma_8 = 0.740$ for the normalization of the linear matter power spectrum, and $n = 0.951$ for the spectral index. The most recent estimates from the Cosmic Microwave Background (CMB) report higher values for both the matter density parameter and $\sigma_8$ \cite{HI13.1,PL13.1}. This however does not change the conclusions of this work.

\section{Preliminary considerations}\label{sct:preliminary}

The SAM developed in this work is based on the halo model \cite{SE00.1,MA00.3,CO02.2}. The halo model is a semi-analytic framework that has been used successfully in order to describe the clustering of galaxies, AGN, various cosmic backgrounds \cite{CO12.1}, and, most importantly for the present discussion, DM. In this latter respect the halo model is based on the assumption that all DM in the Universe is locked inside halos, gravitationally bound structures which can be treated as hard spheres. That being the case, the spatial distribution of DM can be separated into two contributions. The first one consists of particle pairs that belong to the same halo. As such, it depends heavily on the average distribution of DM within halos. This term is dubbed $1-$halo term. The second contribution comes from particle pairs belonging to two separate halos. This term obviously depends on the clustering properties of individual halos, and it is named $2-$halo term. The halo model has the advantage of being physically motivated, and thus it can help to explore regimes not accessible to numerical simulations.

A few additional assumptions for the halo model, specific to this work, need to be spelled out clearly before proceeding. The first one concerns the definition of DM halo in a Universe without baryons. There are different ways to determine the mass and extent of a bound structure, both in observations and in numerical simulations, which makes such a definition ambiguous. In this work I refer to the mass $m$ that is contained within the virial radius $R_\Delta$, defined as the scale encompassing an average density $\Delta=200$ times the average matter density $\bar \rho_\mathrm{m}$. This definition is suggested by compatibility with the most common formulae used for the mass function and halo bias (e.g., \cite{SH01.1,SH02.1}). Under the assumption that all matter is contained within bound structures of some mass, one has that the following relation must hold:

\begin{equation}\label{eqn:constraint1}
\int_0^{+\infty} \mathrm{d}m~m~n(m,z) = \bar\rho_\mathrm{m}~.
\end{equation}
Mass functions $n(m,z)$ calibrated against numerical simulations are usually normalized in a way such that the Eq. (\ref{eqn:constraint1}) is always satisfied.

The situation is more complex when dealing with a universe containing baryons, as well as DM. In this case the average matter density does not coincide with the average DM density, rather it is given by the sum of the DM, gas, and stellar\footnote{The stellar component should in principle also include a contribution from cold gas, which however is expected to be significant only at scales smaller than those relevant to the present work. I hence decided not to include such contribution, although one can easily modify the stellar mass fraction and density profiles to take it into account.} average densities

\begin{equation}
\bar\rho_\mathrm{m} = \bar\rho_\mathrm{DM} + \bar\rho_\mathrm{g}(z) + \bar\rho_\star (z)~.
\end{equation}
Note that the DM density is constant in time, as I am working in comoving coordinates and DM does not get created nor destroyed. On the other hand, gas gets continuosly transformed into stars, and stars revert continuosly into gas, so that their respective comoving densities change with time according to the cosmic star formation history. This difference introduces an additional ambiguity in the definition of a bound structure.

For reasons that will be clear in a moment, I decided to label each structure in a model universe containing baryons by means of the mass $m$ that the same structure would have if evolved in a baryon-free universe. By this I mean a universe with the same primordial density fluctuations but where the initial gas is turned into an equal mass density of additional DM. Henceforth, I shall refer to this mass as the \emph{equivalent mass}. It follows that  each structure with an equivalent mass $m$ placed at redshift $z$ will contain a certain amount $m_\mathrm{DM}(m,z)$ of DM, an amount $m_\mathrm{g}(m,z)$ of hot gas, and an amount $m_\star(m,z)$ of stars. The mass fractions of each matter component can thus be defined as $f_\mathrm{DM}(m,z) \equiv m_\mathrm{DM}(m,z)/m$, $f_\mathrm{g}(m,z) \equiv m_\mathrm{g}(m,z)/m$, and $f_\star(m,z) = m_\star(m,z)/m$, respectively. These definitions allow one to construct the SAM upon the prescriptions for the halo mass function and bias that have been calibrated against $n-$body simulations. For instance, it is possible to write 

\begin{equation}\label{eqn:universalDarkMatterDensity}
\int_0^{+\infty} \mathrm{d}m~mf_\mathrm{DM}(m,z)~n(m,z) = \bar\rho_\mathrm{DM}~,
\end{equation}
where $n(m,z)$ is the standard DM-only halo mass function. This would not have been possible had the \emph{total} mass been used instead of the \emph{equivalent} mass, because the usual mass function prescriptions are not guaranteed to work appropriately when baryons are present \cite{ST09.1,CU12.1,CU13.1}.

This kind of distinction is admittedly unlikely to be important, especially in light of the uncertainties in halo abundances and mass fractions coming from both simulations and observations. However it has the advantage of making the SAM developed here internally self-consistent. Moreover, this approach can easily be applied to the comparison presented in Paper II, because all the simulations presented in the OWLS project share the same initial conditions, and individual structures can easily be matched with their baryon-free counterparts. Likewise, it should be stressed that the sum $f_\mathrm{DM} + f_\mathrm{g} + f_\star$ is in general different from unity, although this difference is usually small. In accordance with the above, the spatial extent of a structure is also considered to coincide with the \emph{equivalent virial radius} $R_\Delta$ that the same structure would have in a baryon-free universe. Mass fraction are always intended to be referred to this spatial extent.

In a way similar to Eq. (\ref{eqn:universalDarkMatterDensity}), the comoving average stellar density can now be written as

\begin{equation}\label{eqn:universalStellarDensity}
\int_0^{+\infty} \mathrm{d}m~mf_\star(m,z)~n(m,z) = \bar\rho_\star(z)~,
\end{equation}
while the treatment of the hot gas requires a little bit more caution. Although the gas distribution at very high redshift has been shown to trace the DM one \cite{GR13.1}, it is well known that at later times a large part of gas in the Universe does not accrete onto DM halos, rather it remains diffused at large scales \cite{BR07.1,SH12.1,GE13.1} (see also \cite{VA13.2} for a recent detection). This incidentally means that the main assumption behind the halo model is not satisfied for this matter component. Later on I will show how this issue has been faced. For the moment I limit myself to observe that if $F_\mathrm{g}(z) < 1$ is the fraction of all gas that at a given redshift $z$ is contained within bound structures, it then follows that

\begin{equation}
\int_0^{+\infty} \mathrm{d}m~mf_\mathrm{g}(m,z)~n(m,z) = F_\mathrm{g}(z)\bar\rho_\mathrm{g}(z)~.
\end{equation}
The total average gas density can be obtained as

\begin{equation}
\bar\rho_\mathrm{g}(z) = \frac{3H_0^2}{8\pi G}\Omega_{\mathrm{b},0} - \bar\rho_\star(z)~,
\end{equation}
where I recall that $\Omega_{\mathrm{b},0}$ is the baryon density parameter.

\section{Mass fractions}

The first and foremost information needed in order to construct a halo model-based SAM is the amount of the different matter components that are included in a structure of a given mass and redshift, i.e., the mass fractions. In this Section I explain how mass fractions have been modeled.

\subsection{Dark Matter}

In order to reduce the amount of free parameters of the SAM I decided to adopt a simplifying assumption and set the DM fraction to a constant,

\begin{equation}
f_\mathrm{DM}(m,z) = 1-\frac{\Omega_{\mathrm{b},0}}{\Omega_{\mathrm{m},0}}~,
\end{equation}
where I remind the reader that the mass fractions are always referred to the equivalent mass, see Section \ref{sct:preliminary}. This assumption is equivalent to reason that the assembly of DM halos is not affected by the presence of baryons. In other words, although an internal rearrangement of the DM distribution is allowed, the total amount of DM that gets accreted by a halo should be invariant with respect to the baryonic physics. In \cite{CU13.1} this is shown to be a fairly good approximation for cosmological simulations of galaxy clusters. The black solid line in Figure \ref{fig:massFractions} shows the DM fraction.

\subsection{Gas}

\begin{figure}
	\centering
	\includegraphics[width=0.70\hsize]{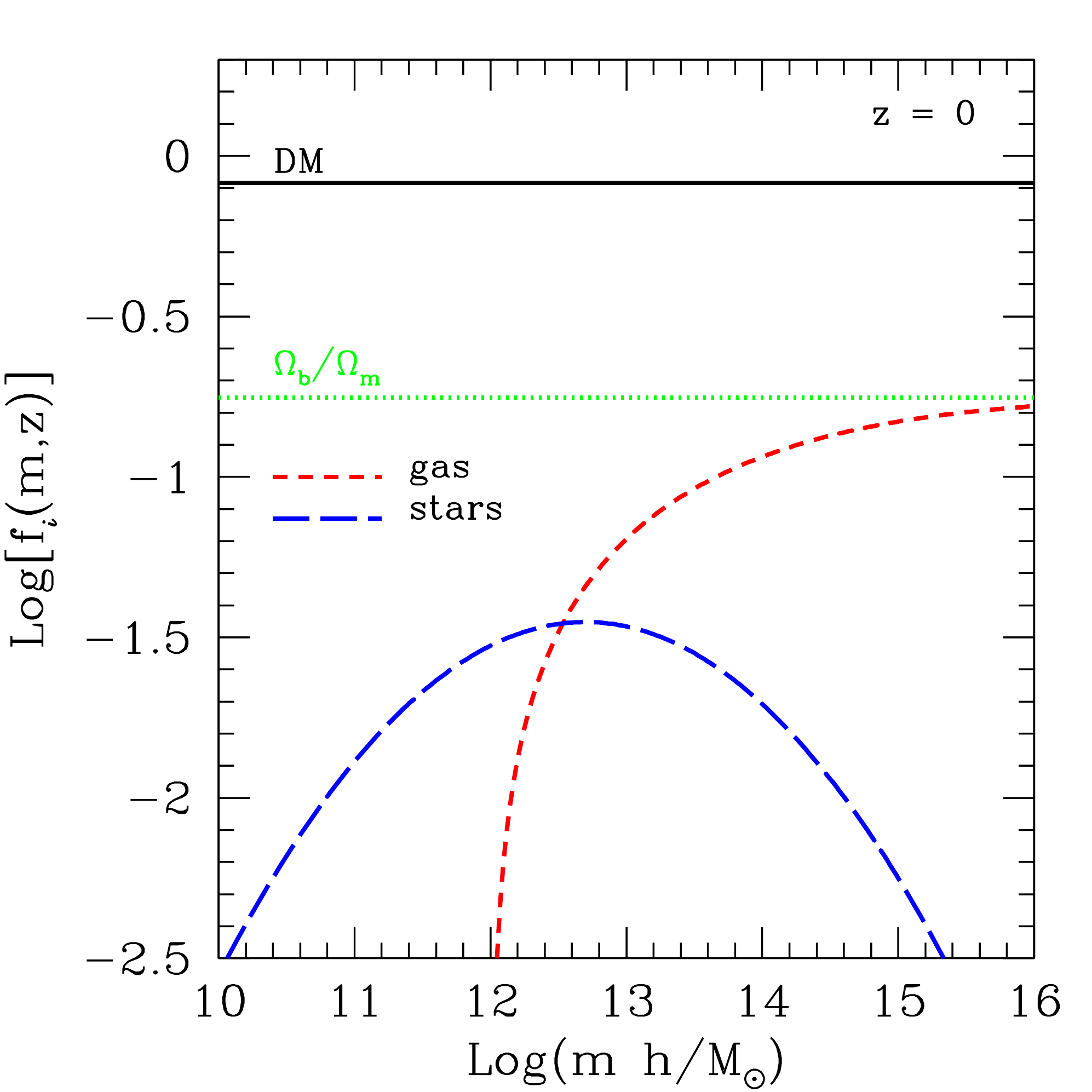}
	\caption{The fractions of DM, hot gas, and stars for a structure with equivalent mass $m$ placed at redshift $z=0$, as labeled. The chosen parameters have been detailed in the text. The horizontal green dotted line represents the universal baryon fraction, to which the gas fraction tends by construction for high masses.}
	\label{fig:massFractions}
\end{figure}

The gas fraction has been studied extensively with X-ray observations of galaxy groups and clusters \cite{VI06.1,GA07.1,GO07.2,GI09.2,PR09.1,SU09.2,DA10.1}, however because of the steep drop in the gas density with radius it is commonly not possible to obtain reliable information at overdensities lower than\footnote{I remind the reader that in this paper overdensities are referred to the \emph{average} matter density of the Universe, so that even an overdensity of $200$ with respect to the \emph{critical} density corresponds to $\Delta = 200/\Omega_{\mathrm{m},0}\sim 840$ at $z=0$.} $\Delta \sim 500$. Extrapolations of the gas content to larger scales are dangerous due to the unknown impact of clumpiness and depletion \cite{EC12.1}. Nevertheless, the generic behavior of the gas fraction as a function of mass can be inferred from these observations and from physical considerations. Specifically, the gas fraction is expected to converge to the universal baryon fraction at large masses. This is due to the fact that $i)$ the mass contribution of stars to massive structures is mostly negligible, and $ii)$ these objects are expected to be little influenced by non-gravitational physics, so that they should be good tracers of cosmological initial conditions. The gas fraction is observed to stay relatively flat or decline slightly with decreasing mass in the group/cluster regime, and then drop sharply at lower masses.

Given these properties, the gas mass fraction has been modeled by a two-parameter error function,

\begin{equation}\label{eqn:gasFraction}
f_\mathrm{g}(m,z) = \frac{\Omega_{\mathrm{b},0}}{\Omega_{\mathrm{m},0}}~\mathrm{erf} \left[ \frac{\mathrm{Log}(m/m_0)}{\sigma} \right]~.
\end{equation}
In Eq. (\ref{eqn:gasFraction}) the parameter $m_0$ defines the position of the drop in gas fraction, while the parameter $\sigma$ defines its sharpness. Note that this function becomes negative for $m < m_0$. I simply set $f_\mathrm{g}=0$ in that case. I did not include any explicit dependence of the gas fraction on redshift, thus assuming that such dependence is implicit in the two free parameters. The red short-dashed line in Figure \ref{fig:massFractions} shows the gas fraction calculated according to Eq. (\ref{eqn:gasFraction}) by adopting $m_0=10^{12} h^{-1}M_\odot$ and $\sigma=3$. These parameter values have been chosen for the purpose of illustration only. One can choose his or her preferred gas mass fraction relation and fit the parameters $m_0$ and $\sigma$ accordingly. The latter procedure has been followed in Paper II.

\subsection{Stars}

The stellar fraction has also been subject of considerable observational attention \citep{BA07.1,CO07.1,GO07.2,GI09.2,AN10.1,BE10.1,LE12.1,GO13.1,BU13.1}. In this case the results are also often limited to large overdensities, and suffer from substantial uncertainties linked with the contribution of small satellites, the intra-halo light, and the assumed models for converting stellar magnitudes into masses (see \cite{GO13.1} for a thorough discussion of these systematics). Generically, the stellar fraction becomes small in massive clusters, where the gas dominates the baryon budget, and it is expected to vanish for small objects, since their potential wells are not deep enough to keep gas at a density high enough to allow cooling and star formation. Star formation should peak at somewhat intermediate masses, where the peak mass ranges between $\sim 5\times 10^{11}h^{-1}M_\odot$ and $\sim 10^{13}h^{-1}M_\odot$, depending on the study one refers to.

In accordance with this, the mass-dependence of the stellar fraction has been represented by a Gaussian,

\begin{equation}
f_\star (m,z) = A\exp\left[ -\frac{\mathrm{Log}^2(m/m_0)}{2\sigma^2} \right].
\end{equation}
In order to reduce the number of free parameters, the amplitude $A$ of the Gaussian has been fixed by requesting that the universal stellar density (Eq. \ref{eqn:universalStellarDensity}) matches a certain value. When applying the SAM to cosmological simulations (in Paper II) this value is the stellar density measured in the simulation at hand. In this case, for the purpose of illustration, I considered the value $7\times 10^8 h^2 M_\odot$ Mpc$^{-3}$, which agrees reasonably well with observations at $z=0$ \cite{RU03.1,BE07.3,CO08.1,WI08.2,LU12.1}. If one is interested in different redshifts, then he or she has to select a value of the stellar density that is appropriate for that redshift. This procedure makes sense as the paucity of observational data is such that in many circumstances three free parameters are too many to obtain a reliable fit. Once more, the redshift dependence is implicit through the universal stellar mass density and the two independent parameters $m_0$ and $\sigma$. The blue long-dashed line in Figure \ref{fig:massFractions} shows the stellar fraction by assuming $m_0=5\times 10^{12} h^{-1}M_\odot$ and $\sigma=1.2$. I stress that this choice of parameters is illustrative only. The slope of the stellar fraction as a function of group and cluster masses is still a matter of substantial debate, with recent works finding it to be significantly shallower than shown in Figure \ref{fig:massFractions} \cite{BU13.1} (see however \cite{GO13.1}).

\section{Density profiles}

The next elements of the halo model SAM that need to be specified are the density runs $\rho_i(r|m)$ of the different matter components within a given structure. Note that these profiles in general depend on the equivalent mass and redshift of the structure. However, for the sake of clarity, from now on I omitted the redshift dependence, except where ambiguity can arise. For future use, let me also introduce the Fourier transform of the density profiles, as

\begin{equation}\label{eqn:fourier}
\hat\rho_i(k|m) = 4\pi R_\Delta^3 \int_0^1 x^2\mathrm{d}x~\rho_i(R_\Delta x|m)~j_0(kR_\Delta x)~,
\end{equation}
where $j_0(t)$ is the spherical Bessel function of order zero. Note that the Fourier integral is truncated at the equivalent virial radius, so that in the limit $kR_\Delta \rightarrow 0$ one obtains $\hat\rho_i(k|m) \rightarrow mf_i(m)$. In the SAM each density profile is characterized by a certain number of parameters. Each individual parameter $p$ is allowed to have a power-law dependence on the equivalent structure mass, that is

\begin{equation}\label{eqn:parameter}
p(m) = p_0\left( \frac{m}{10^{12}h^{-1}M_\odot} \right)^\theta~.
\end{equation}
Similarly to the mass fractions, also in this case the redshift dependence is implicitly introduced through the power-law parameters.

It is worth stressing that this formalism assumes universal shapes for the density profiles of the various matter components (only the parameters change with mass). While this is an excellent approximation for DM, it might not be so for gas and stars, for which non-gravitational physical processes are likely to introduce characteristic scales and thus break the universality of the profiles. For this first version of the SAM I stick to this assumption, that will be shown to work rather well in Paper II. It will be possible to relax such an assumption in subsequent versions.

\subsection{Dark matter}

DM density profiles are well known to follow a universal or quasi-universal functional form that is valid for a wide range of masses and scales \cite{NA96.1,NA97.1,GA08.1,NA04.1}. This universal function has been introduced for the first time by Navarro, Frenk, \& White \cite{NA95.1}, and is referred to as NFW henceforth. It is the functional form that I adopted to describe the DM average density, and can be written as

\begin{equation}
\rho_\mathrm{DM}(x|m) = \frac{\rho_\mathrm{s}}{x(1+x)^2}~,
\end{equation}
where $x \equiv r/r_\mathrm{s}$. The two parameters $\rho_\mathrm{s}$ and $r_\mathrm{s}$ are the scale density and scale radius of the profile, respectively. Because there is a correlation between the formation time of a DM halo and its degree of compactness \cite{NA96.1,BU01.1,EK01.1}, the two parameters of the NFW profile depend on each other, and can thus be reduced to only one. By defining the concentration as $c\equiv R_\Delta/r_\mathrm{s}$, the scale density is constrained by the requirement that the total DM mass within the structure matches the DM mass fraction:

\begin{equation}
\rho_\mathrm{s} = f_\mathrm{DM}\frac{\Delta}{3}\bar\rho_\mathrm{m}\frac{c^3}{G(c)}~,
\end{equation}
where $G(c)$ is the usual NFW concentration function.

Because of the correlation mentioned above, the concentration can be linked to the equivalent mass. There are a number of prescriptions for this correlation  in the literature \cite{NA96.1,BU01.1,EK01.1,DO04.1,DU08.1,GA08.1,PR12.1}, all derived from the analysis of $n-$body simulations. The presence of baryons tends to modify DM concentrations \cite{GN04.1,DU10.1}, in ways that depend on the exact physical processes that are being considered. More details on this are given in what follows and in Paper II. For the time being I just notice that, if one is interested only in the matter clustering \emph{relative} to the DM clustering in the baryon-free scenario, then the specific concentration-mass relation adopted in the latter is only marginally relevant. What truly is important is only the \emph{relative} change in DM concentrations due to non-gravitational physics. For this reason I set the only free parameter of the DM profile to be $p=c/c_\mathrm{DMONLY}$, where $c_\mathrm{DMONLY}$ is the reference concentration in the baryon-free case. For this reference concentration-mass relation I adopted the prescription

\begin{equation}
c_\mathrm{DMONLY}(m) = 11~\left(\frac{m}{10^{12}h^{-1}M_\odot}\right)^{-0.1}~,
\end{equation}
which is very similar to the prescription in \cite{DU08.1}. Note that by setting $p_0=1$ and $\theta=0$ in Eq. (\ref{eqn:parameter}) above one recovers $c=c_\mathrm{DMONLY}$.

\begin{figure}
	\centering
	\includegraphics[width=0.49\hsize]{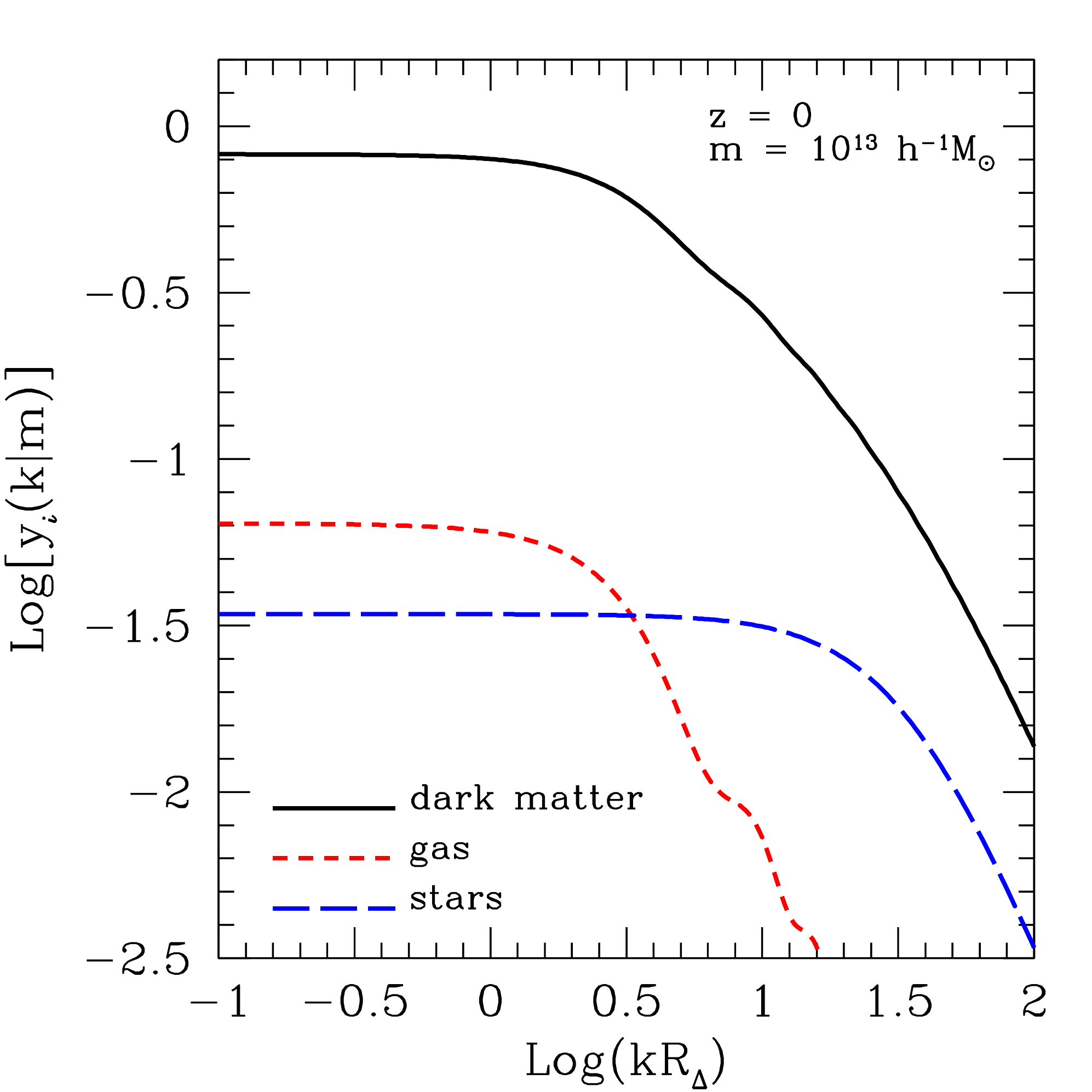}
	\includegraphics[width=0.49\hsize]{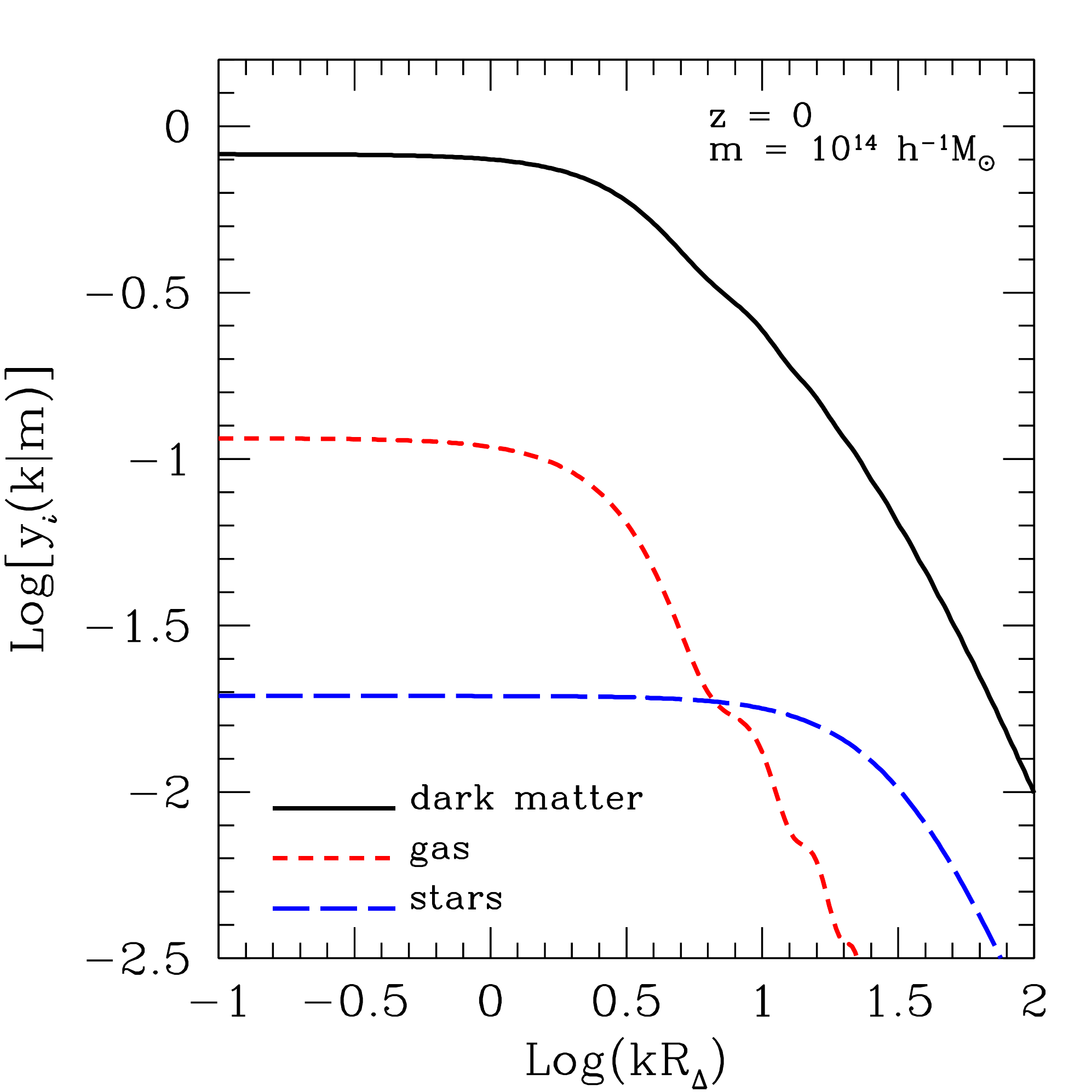}
	\caption{The Fourier transforms of the density profiles of stars, hot gas, and DM, as labeled. The left panel refers to a structure with equivalent mass $m=10^{13}h^{-1}M_\odot$ placed at redshift $z=0$, while in the right panel $m=10^{14}h^{-1}M_\odot$ at the same redshift.}
	\label{fig:profiles}
\end{figure}

Irrespective of how its concentration gets estimated, the Fourier transform of the NFW functional form can be computed analytically \cite{SC01.1} as

\begin{equation}
y_\mathrm{DM}(k|m)\equiv\frac{\hat\rho_\mathrm{DM}(k|m)}{m} = 4\pi\frac{\rho_\mathrm{s}r_\mathrm{s}^3}{m} \left[ \sin(kr_\mathrm{s})\gamma_\mathrm{s}(k) + \cos(kr_\mathrm{s})\gamma_\mathrm{c}(k) - \frac{\sin(ckr_\mathrm{s})}{(1+c)kr_\mathrm{s}} \right]~,
\end{equation}
where

\begin{equation}
\gamma_\mathrm{s}(k) = \mathrm{Si}[(1+c)kr_\mathrm{s}] - \mathrm{Si}(kr_\mathrm{s})
\end{equation}
and 

\begin{equation}
\gamma_\mathrm{c}(k) = \mathrm{Ci}[(1+c)kr_\mathrm{s}] - \mathrm{Ci}(kr_\mathrm{s})~.
\end{equation}
The functions Si$(t)$ and Ci$(t)$ are the sine integral and cosine integral, respectively. Note that as $k \rightarrow 0$, $y_\mathrm{DM}(k|m) \rightarrow f_\mathrm{DM}(m)$ as expected. The black solid lines in Figure \ref{fig:profiles} represent the Fourier transforms of the DM density profiles for two halos with masses $m=10^{13}h^{-1}M_\odot$ and $m=10^{14}h^{-1}M_\odot$, respectively, by assuming $\theta = 0$ and $p_0 = 2$ in order to mimic baryon contraction (see below).

\subsection{Gas}

The gas distribution has been modeled by a $\beta-$model \cite{CA76.1},

\begin{equation}
\rho_\mathrm{g}(r|m) = \frac{\rho_\mathrm{c}}{\left( 1+x^2 \right)^{3\beta/2}}~,
\end{equation}
where in this case $x\equiv r/r_\mathrm{c}$. The $\beta-$model is a cored profile, where the characteristic size of the core is given by the parameter $r_\mathrm{c}$. The parameter $\beta$ defines instead the outer slope of the gas distribution. In a similar fashion to the DM profile, the core density $\rho_\mathrm{c}$ of the gas profile is selected by requiring that the gas mass fraction is matched. This means that the following relation is enforced:

\begin{equation}
\frac{4}{3}\pi\rho_\mathrm{c}r_\mathrm{c}^3 \left[ x_\Delta^3\phantom{|}_2F_1\left( \frac{3}{2}, \frac{3}{2}\beta, \frac{5}{2};-x_\Delta^2\right) \right] = mf_\mathrm{g}(m)~.
\end{equation}
In the previous Equation $x_\Delta \equiv R_\Delta/r_\mathrm{c}$, and $\phantom{|}_2F_1(a,b,c;t)$ is the Gauss hypergeometric function.

Unfortunately, the Fourier transform of a $\beta-$model cannot be computed analytically, except in the very specific case in which $\beta = 2/3$. For the sake of generality, the SAM always computes the Fourier transform of the gas profile numerically. The red short-dashed lines in Figure \ref{fig:profiles} show the Fourier transforms of the gas density profiles for two structures with masses $m=10^{13}h^{-1}M_\odot$ and $m=10^{14}h^{-1}M_\odot$, respectively. For the purpose of illustration I did not include any explicit mass dependence for the gas profile parameters, and simply set $\beta=2/3$ and $r_\mathrm{c}/R_\Delta = 0.05$ throughout. These values roughly match observations of X-ray galaxy clusters \cite{MA98.1,NE99.1,OT04.1,VI06.1,CR08.2,CA10.2}.

\subsection{Stars}

According to observations of galaxy groups and clusters, the stellar distribution within DM halos is fairly well represented by a NFW profile with a low concentration \cite{LI04.1,GO13.1}. However on galaxy scales it is expected that the stellar profile be substantially more compact than an NFW (e.g., \cite{KO06.1}). Given that the stellar cusp at the center of individual galaxies is at too small scales to be relevant for the present study, I chose to adopt a profile with the same inner slope of the NFW, suitable for group-scale and cluster-scale structures\footnote{Recent results from the CLASH team (see for instance \cite{CO12.2}) show that the overall potential wells of galaxy clusters are also very well fit by a NFW profile down to very small scales, lending support to the approach used here.}, with the addition of an exponential drop (modulated by an exponent $\alpha$) which allows to mimic the sharper drop of lower-mass structures. In Paper II we found that this approximation is sufficiently accurate to describe the clustering of stars in the OWLS simulations, however more accurate prescriptions can be easily implemented in the future. The resulting stellar density profile is

\begin{equation}
\rho_\star (x|m) = \frac{\rho_\mathrm{t}}{x}\exp(-x^\alpha)~,
\end{equation}
with $x\equiv r/r_\mathrm{t}$. The parameter $r_\mathrm{t}$ defines the transition between the inner cusp and the outer exponential decline of the profile. Furthermore, if $\alpha = 1$ one obtains the standard exponential disk profile.

As with the gas profile, the transition density $\rho_\mathrm{t}$ has been set in order to reproduce the required stellar fraction in the structure at hand. This means that the following relation

\begin{equation}
\frac{4\pi}{\alpha} \rho_\mathrm{t}r_\mathrm{t}^3\left[ \Gamma\left(1-\nu(\alpha)\right)-x_\Delta^2E_{\nu(\alpha)}(x_\Delta^\alpha)\right] = mf_\star(m)~,
\end{equation}
with $x_\Delta=R_\Delta/r_\mathrm{t}$, has to be satisfied. In the previous Equation $E_\nu(t)$ is the exponential integral of order $\nu$, and

\begin{equation}
\nu(\alpha) \equiv 1-\frac{2}{\alpha}~.
\end{equation}
The Fourier transform of this profile can be obtained analytically only for the special case in which $\alpha = 1$, and it has in general to be evaluated numerically. The blue long-dashed curves in Figure \ref{fig:profiles} show the Fourier transforms of the stellar density profiles for two structures with masses $m=10^{13} h^{-1}M_\odot$ and $m=10^{14}h^{-1}M_\odot$, respectively. Similarly to the gas, here I did not include any explicit mass dependence, and I just set $\alpha = 1$ and  $r_\mathrm{t}/R_\Delta = 0.03$ \cite{GU10.1,LU11.1}.

\subsection{Implications}

Figure \ref{fig:profiles} compares the Fourier transforms of the various mass density profiles within two structures of different masses. The fact that the stellar density profile is much more compact than its gas and DM counterparts is reflected by the fact that its Fourier transform has a substantially wider core at small wavenumbers. For the same reason, the Fourier transform of the gas density profile is significantly narrower than the others. As expected, the large-scale plateaus in Fourier space coincide with the relative abundances of the different mass components. Looking more closely at Figure \ref{fig:profiles}, oscillations are visible in the Fourier transforms of the DM and gas density runs. These are due to the fact that the Fourier transforms are truncated at the equivalent virial radius (Eq. \ref{eqn:fourier}). Oscillations in the gas transform are stronger because the gas density profile is flatter, and thus the truncation is sharper, than for the other profiles. With information of the mass fractions and matter distributions it is now possible to use the halo model to write down the power spectra of the individual components.

\section{Matter power spectra}

\subsection{Dark matter}

The DM power spectrum can be computed through the standard implementation of the halo model sketched in Section \ref{sct:preliminary} (see \cite{SE00.1,MA00.3,CO02.2} for more details). Accordingly, the power spectrum is given by the sum of the $1-$halo and $2-$halo terms as $P_\mathrm{DM}(k) = P_\mathrm{DM}^{(1)}(k)+P_\mathrm{DM}^{(2)}(k)$, where

\begin{equation}\label{eqn:darkMatter1halo}
P_\mathrm{DM}^{(1)}(k) = \frac{1}{\bar\rho^2_\mathrm{DM}}\int_0^{+\infty} \mathrm{d}m~n(m)~m^2y_\mathrm{DM}^2(k|m)
\end{equation}
and

\begin{equation}\label{eqn:darkMatter2halo}
P_\mathrm{DM}^{(2)}(k) = \frac{P_\mathrm{L}(k)}{\bar\rho_\mathrm{DM}^2}\left[ \int_0^{+\infty} \mathrm{d}m~n(m)~b(m)~my_\mathrm{DM}(k|m) \right]^2~,
\end{equation}
respectively. In the previous Eq. (\ref{eqn:darkMatter2halo}) the function $b(m)$ represents the bias of DM halos, while $P_\mathrm{L}(k)$ is the linear DM power spectrum. It is physically meaningful that on large scales the non-linear DM power spectrum has to converge to the linear one. This means that $P_\mathrm{DM}^{(2)}(k)\rightarrow P_\mathrm{L}(k)$ for $k\rightarrow 0$, and hence the following non-trivial condition has to be enforced:

\begin{equation}\label{eqn:constraint}
\int_0^{+\infty} \mathrm{d}m~n(m)~b(m)~mf_\mathrm{DM}(m) = \bar\rho_\mathrm{DM}~.
\end{equation}
I imposed this condition by adding a constant to the bias function in the smallest mass bin used in the numerical integration, as detailed in \cite{RE02.2,AM04.1,FE10.1}.

The (dark) matter power spectrum $P_\mathrm{DMONLY}(k)$ in the baryon-free case can also be obtained through Eqs. (\ref{eqn:darkMatter1halo}) and (\ref{eqn:darkMatter2halo}) by setting $f_\mathrm{DM} (m) = 1$. In the remainder of this paper I considered only relative deviations from $P_\mathrm{DMONLY}(k)$, thus replacing the power spectra $P_i(k)$ of the various matter components and their cross spectra $P_{ij}(k)$ with the dimensionless functions $U_i(k)\equiv P_i(k)/P_\mathrm{DMONLY}(k)$ and $U_{ij}(k)\equiv P_{ij}(k)/P_\mathrm{DMONLY}(k)$, respectively.

\begin{figure}
	\centering
	\includegraphics[width=0.70\hsize]{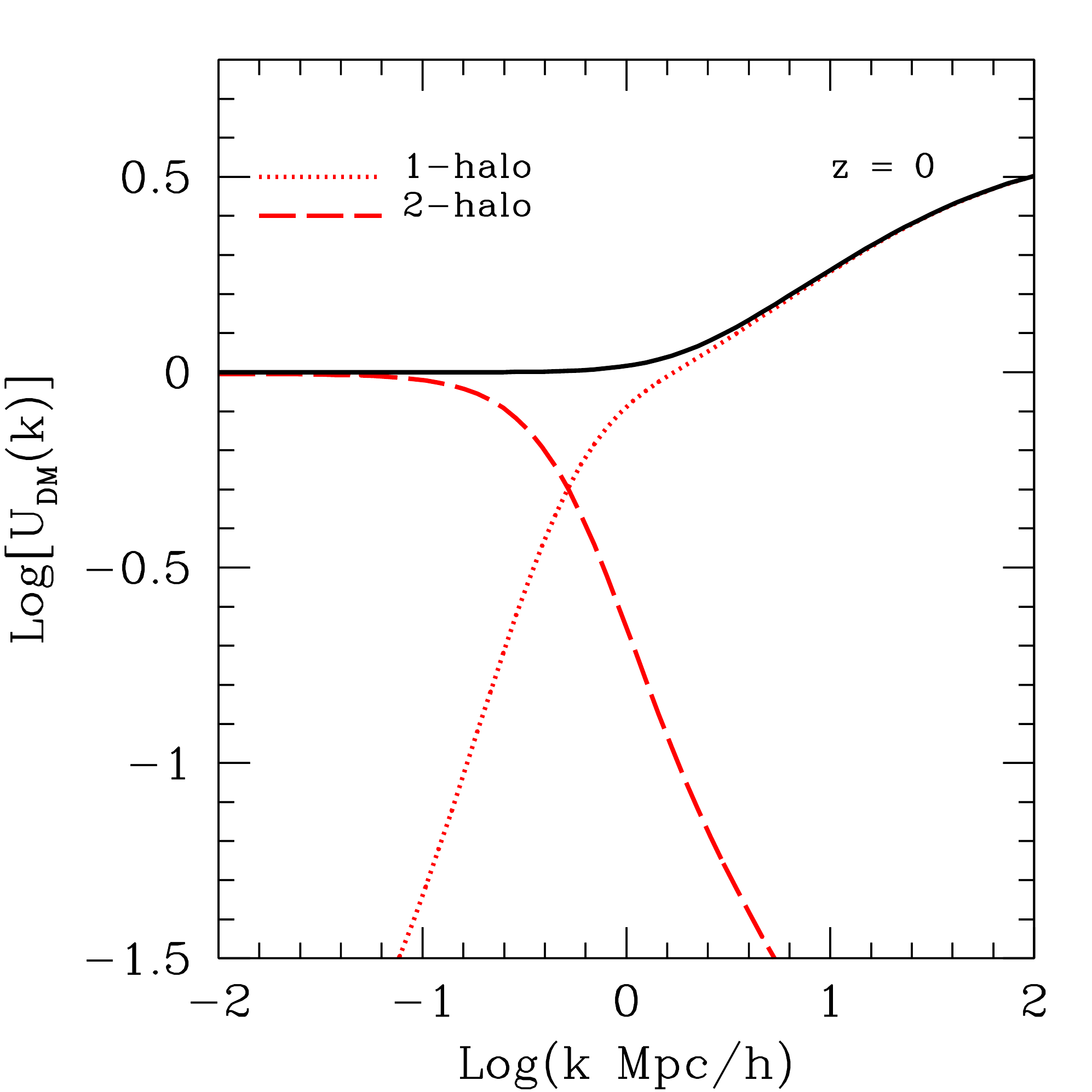}
	\caption{The DM power spectrum as a function of scale, for $z=0$. In order to illustrate the effect of baryon physics, the concentration of DM halos has been increased by a factor of $2$, irrespectively of the mass, with respect to the baryon-free case. The two red lines represent the $1-$halo and $2-$halo contributions, as labeled.}
	\label{fig:darkMatterPower}
\end{figure}

In Figure \ref{fig:darkMatterPower} I show the DM power spectrum, as well as its two components. In order to illustratively mimick the effect of non-gravitational physics I increased the DM halo concentrations by a factor of $2$, irrespective of the mass, with respect to the baryon-free case. Such a modification causes the increment in power that is observed for $k \gtrsim 1 h$ Mpc$^{-1}$. Increasing the concentration of DM halos is consistent with the fact that gas cooling and star formation deepen the potential well of a structure, thus causing the DM profile to contract \cite{BL86.1,GN04.1}. Several hydrodynamical simulations of structure formation do indeed show an increment in DM power at small scales, compatible with this scenario \cite{RU08.2,FE11.1}. However realistic situations are more complex than a simple constant increment in DM concentrations. Baryon contraction is not expected to work (or to work at the same level) for all halo masses. Moreover, the presence of strong energy feedback can halt, and possibly reverse, the contraction \cite{DU10.1,MC10.1,MC11.1,VA11.1}. This issue is explored in full detail in Paper II, making use of the OWLS set.

\subsection{Gas}

The situation is substantially more complicated when dealing with gas. Indeed, the halo model in its standard form can be used only for the gas that is bound to DM halos. However this is only the minority of the gas in the Universe, while the majority is spread outside bound structures. I refer to this latter component as to \emph{diffuse} gas. Because the gas is divided in two components, the gas power spectrum can be divided into four distinct contributions: $i)$ the contribution from particle pairs sitting in the same structure ($1-$halo); $ii)$ the contribution from particle pairs residing in two distinct structures ($2-$halo); $iii)$ the contribution from particle pairs both belonging to the diffuse component; $iv)$ the contribution from particle pairs where one particle is bound to a structure and the other belongs to the diffuse component. By working out the clustering statistics it is easy to see that the gas power spectrum can be written as

\begin{equation}\label{eqn:gasPower}
P_\mathrm{g}(k) = \left( 1-F_\mathrm{g} \right)^2P_\mathrm{g,d}(k)+ 2F_\mathrm{g}\left( 1-F_\mathrm{g} \right)P_\mathrm{g,dh}(k)+F_\mathrm{g}^2\left[P_\mathrm{g,h}^{(1)}(k)+P_\mathrm{g,h}^{(2)}(k)\right]~.
\end{equation}
The $1-$halo and $2-$halo contributions appearing in the rightmost term of the previous Equation can be written analogously to Eqs. (\ref{eqn:darkMatter1halo}) and (\ref{eqn:darkMatter2halo}) for the DM case,

\begin{equation}
P_\mathrm{g,h}^{(1)}(k) = \frac{1}{F^2_\mathrm{g}\bar\rho^2_\mathrm{g}} \int_0^{+\infty} \mathrm{d}m~n(m)~m^2y_\mathrm{g}^2(k|m)~
\end{equation}
and

\begin{equation}
P_\mathrm{g,h}^{(2)}(k) = \frac{P_\mathrm{L}(k)}{F_\mathrm{g}^2\bar\rho^2_\mathrm{g}} \left[ \int_0^{+\infty} \mathrm{d}m~n(m)~b(m)~my_\mathrm{g}(k|m)\right]^2~.
\end{equation}

For the treatment of the smooth component I followed a recent work by Smith \& Markovic \cite{SM11.1} on the clustering of warm DM. In that work the authors were also faced with the problem of using the halo model to compute the power spectrum of a matter component with a substantial diffuse constituent (warm DM rather than gas). In the present case, by assuming that the density field of the diffuse gas component is related to the total mass density field by a deterministic local mapping, then at first (linear) order the power spectrum of this diffuse gas is equal to the linear DM power spectrum times a constant (albeit redshift-dependent) bias. Namely,

\begin{equation}
P_\mathrm{g,d}(k) = b^2_\mathrm{d}~P_\mathrm{L}(k)~.
\end{equation}
Smith \& Markovic used an argument based on DM mass conservation in order to infer the value of the diffuse bias. Here the same argument cannot be applied because gas is a matter component distinct from DM. In Paper II it will be shown that the value of $b_\mathrm{d}$ depends on the non-gravitational physics that one is considering, as stronger feedback implies a bias closer to unity. In any case the value of the bias turns out to be always $b_\mathrm{d} \lesssim 1$ \cite{JI06.1,RU08.1,FE11.1}, thus for the time being I illustrated the results obtained with the value $b_\mathrm{d}=0.85$. On a related note, it is important to emphasize that Van Waerbeke, Hinshaw, and Murray \cite{VA13.2} have recently cross-correlated cosmic shear with thermal Sunyaev-Zel'dovich maps and convincingly detected for the first time the presence of this diffuse gas component. It also turns out that a constant bias factor describes well its clustering properties, although their measured bias depends upon the gas temperature and density. By setting $b_\mathrm{d}=0.85$ returns a temperature for the diffuse gas component of $\sim 2$ keV and an electron number density of $\sim 1$ m$^{-3}$.

Following \cite{SM11.1} the cross-spectrum between the diffuse component and the halo component can easily be written as

\begin{equation}
P_\mathrm{g,dh}(k) = b_\mathrm{d}\frac{P_\mathrm{L}(k)}{F_\mathrm{g}\bar\rho_\mathrm{g}} \int_0^{+\infty} \mathrm{d}m~n(m)~b(m)~my_\mathrm{g}(k|m)~.
\end{equation}
It should be noted that in the limit of very large scales, the gas power spectrum approaches

\begin{equation}
P_\mathrm{g}(k)\rightarrow P_\mathrm{L}(k)\left[ \left( 1-F_\mathrm{g} \right)^2b^2_\mathrm{d} + 2 F_\mathrm{g} \left( 1- F_\mathrm{g}\right) b_\mathrm{d}b_\mathrm{g,eff} + F_\mathrm{g}^2b_\mathrm{g,eff}^2 \right]\simeq b_\mathrm{d}^2\;P_\mathrm{L}(k)~,
\end{equation}
where $b_\mathrm{g,eff}$ is the effective bias of gas contained within bound structures. The last approximate equality follows because in general $F_\mathrm{g} \ll 1$.

\begin{figure}
	\centering
	\includegraphics[width=0.70\hsize]{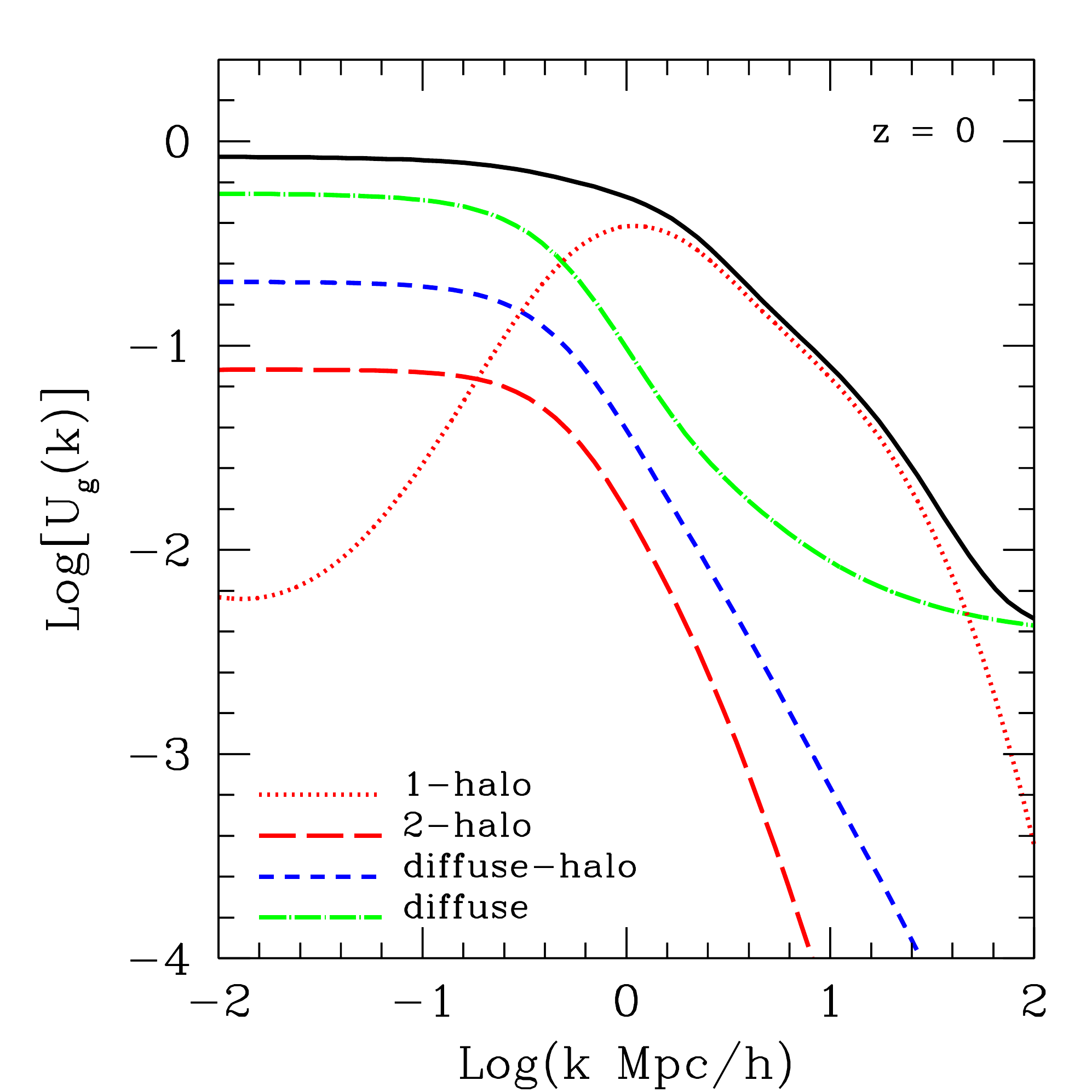}
	\caption{The gas power spectrum as a function of scale, for $z=0$. The four colored lines represent separate contributions to the total spectrum, as labeled. Each contribution is weighted by the appropriate prefactor, as in Eq. (\ref{eqn:gasPower}), so that the total spectrum (black solid line) is simply the sum of the four.}
	\label{fig:gasPower}
\end{figure}

In Figure \ref{fig:gasPower} I show the total gas power spectrum according to the SAM, separated into its four components. Note that each component is weighted by the appropriate prefactor, as in Eq. (\ref{eqn:gasPower}), so that the total spectrum is simply the sum of the four. This Figure contains some interesting information. Firstly, on large scales ($k \lesssim 1 h$ Mpc$^{-1}$) the dominant contribution to the gas power spectrum is given by the diffuse gas component. In this regime gas density fluctuations induced by the clustering of bound structures contribute less than $10\%$ to the total gas power. This implies that detection of clustered gas on these scales (e.g., \cite{VA13.2}) refers almost entirely to the diffuse component. Secondly, for wavenumbers larger than $k\sim 1h$ Mpc$^{-1}$ the gas clustering signal is dominated by the $1-$halo term, so that its strength has to depend substantially on the gas mass fraction and average density profile in galaxy groups and clusters. On the other hand, the contributions of the $2-$halo term and of the diffuse-halo cross spectrum are always subdominant. The SAM presented here could thus also be useful for the interpretation of low-density Warm-Hot Intregalactic Medium (WHIM) observations.

\subsection{Stars}

\begin{figure}
	\centering
	\includegraphics[width=0.70\hsize]{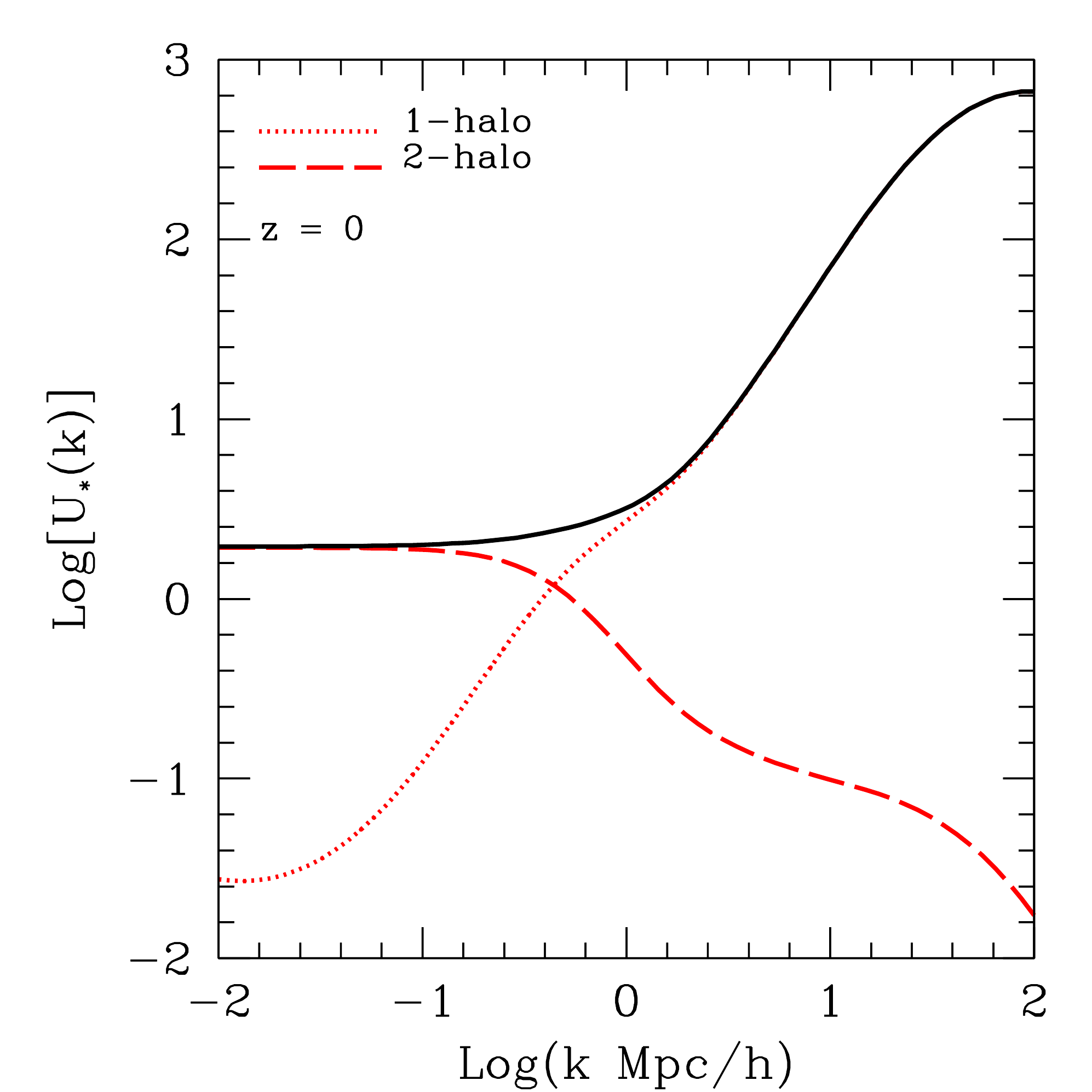}
	\caption{The stellar power spectrum as a function of scale, for $z=0$. The two red lines represent the $1-$halo and $2-$halo contributions, as labeled.}
	\label{fig:stellarPower}
\end{figure}

Because stars can be found only within bound structures, the stellar power spectrum can be written in analogy with the DM one, $P_\star(k) = P_\star^{(1)}(k)+P_\star^{(2)}(k)$, where

\begin{equation}
P_\star^{(1)}(k) = \frac{1}{\bar\rho^2_\star} \int_0^{+\infty} \mathrm{d}m~n(m)~m^2y_\star^2(k|m)~
\end{equation}
and

\begin{equation}
P_\star^{(2)}(k) = \frac{P_\mathrm{L}(k)}{\bar\rho^2_\star} \left[ \int_0^{+\infty} \mathrm{d}m~n(m)~b(m)~my_\star(k|m)\right]^2~,
\end{equation}
respectively.

Note that, because of Eq. (\ref{eqn:universalStellarDensity}), as $k\rightarrow 0$ one has that $P_\star^{(2)}(k) \rightarrow P_\mathrm{L}(k)b^2_{\star,\mathrm{eff}}$, where $b_{\star,\mathrm{eff}}$ is the effective bias of stellar clumps. In Figure \ref{fig:stellarPower} I display the stellar power spectrum according to the SAM, as well as its two separate contributions. As can be seen, the stellar power is always higher than the DM power in the baryon-free case. At large scales this is due to the fact that stars form in halos that are biased with respect to the underlying DM distribution. At small scales, this is due to the fact that stellar profiles are much more compact than DM profiles.

\subsection{Cross-spectra}

In addition to the power spectra of the three matter components, their mutual cross spectra need also to be considered in order to compute the total matter power spectrum. In this Section I give formulae for evaluating each one of the cross spectra according to the halo model SAM, which can easily be derived from the clustering statistics \cite{CO02.2}. However, for the sake of brevity, I do not show individual plots. The relative contributions of the individual power spectra and cross spectra to the total matter power spectrum can be appreciated in Figure \ref{fig:matterPower} below.

The cross-correlation between DM and stars is the easiest one to evaluate, since both the involved mass components reside only within bound structures. It follows that their cross spectrum can be written as the sum of a $1-$halo contribution and a $2-$halo contribution: $P_{\mathrm{DM}\star}(k) = P_{\mathrm{DM}\star}^{(1)}(k) + P_{\mathrm{DM}\star}^{(2)}(k)$, where

\begin{equation}
P^{(1)}_{\mathrm{DM}\star}(k) = \frac{1}{\bar\rho_\mathrm{DM}\bar\rho_\star}\int_0^{+\infty} \mathrm{d}m~n(m)~m^2y_\mathrm{DM}(k|m)y_\star(k|m)
\end{equation}
and 

\begin{equation}
P^{(2)}_{\mathrm{DM}\star}(k) = \frac{P_\mathrm{L}(k)}{\bar\rho_\mathrm{DM}\bar\rho_\star}\left[ \int_0^{+\infty} \mathrm{d}m~n(m)~b(m)~my_\mathrm{DM}(k|m)\right]\left[ \int_0^{+\infty} \mathrm{d}m~n(m)~b(m)~my_\star(k|m)\right],
\end{equation}
respectively.

The cross-correlation between DM and gas is more complex to tractate. Because gas has a diffuse component, three separate contributions can be identified: $i)$ pairs made by particles that belong to the same structure ($1-$halo); $ii)$ pairs made by particles belonging to separate structures ($2-$halo); $iii)$ pairs made by gas particles belonging to the diffuse component. It easily follows that the DM-gas cross spectrum takes the form

\begin{equation}
P_{\mathrm{DM}\mathrm{g}}(k) = \left( 1-F_\mathrm{g} \right)P_{\mathrm{DM}\mathrm{g,dh}}(k) + F_\mathrm{g}\left[ P^{(1)}_{\mathrm{DM}\mathrm{g,h}}(k) + P^{(2)}_{\mathrm{DM}\mathrm{g,h}}(k) \right]~,
\end{equation}
where 

\begin{equation}
P_{\mathrm{DM}\mathrm{g,dh}}(k) = b_\mathrm{d}\frac{P_\mathrm{L}(k)}{\bar\rho_\mathrm{DM}} \int_0^{+\infty} \mathrm{d}m~n(m)~b(m)~my_\mathrm{DM}(k|m)~,
\end{equation}

\begin{equation}
P^{(1)}_{\mathrm{DM}\mathrm{g,h}}(k) = \frac{1}{\bar\rho_\mathrm{DM} F_\mathrm{g}\bar\rho_\mathrm{g}}\int_0^{+\infty} \mathrm{d}m~n(m)~m^2y_\mathrm{DM}(k|m)y_\mathrm{g}(k|m)
\end{equation}
and 

\begin{equation}
P^{(2)}_{\mathrm{DM}\mathrm{g,h}}(k) = \frac{P_\mathrm{L}(k)}{\bar\rho_\mathrm{DM} F_\mathrm{g}\bar\rho_\mathrm{g}}\left[ \int_0^{+\infty} \mathrm{d}m~n(m)~b(m)~my_\mathrm{DM}(k|m)\right]\left[ \int_0^{+\infty} \mathrm{d}m~n(m)~b(m)~my_\mathrm{g}(k|m)\right],
\end{equation}
respectively.

Using a similar reasoning, the cross spectrum of gas and stars can be written as

\begin{equation}
P_{\star\mathrm{g}}(k) = \left( 1-F_\mathrm{g} \right)P_{\star\mathrm{g,sh}}(k) + F_\mathrm{g}\left[ P^{(1)}_{\star\mathrm{g,h}}(k) + P^{(2)}_{\star\mathrm{g,h}}(k) \right]~,
\end{equation}
where 

\begin{equation}
P_{\star\mathrm{g,sh}}(k) = b_\mathrm{d}\frac{P_\mathrm{L}(k)}{\bar\rho_\star} \int_0^{+\infty} \mathrm{d}m~n(m)~b(m)~my_\star(k|m)~,
\end{equation}

\begin{equation}
P^{(1)}_{\star\mathrm{g,h}}(k) = \frac{1}{\bar\rho_\star F_\mathrm{g}\bar\rho_\mathrm{g}}\int_0^{+\infty} \mathrm{d}m~n(m)~m^2y_\star(k|m)y_\mathrm{g}(k|m)
\end{equation}
and 

\begin{equation}
P^{(2)}_{\star\mathrm{g,h}}(k) = \frac{P_\mathrm{L}(k)}{\bar\rho_\star F_\mathrm{g}\bar\rho_\mathrm{g}}\left[ \int_0^{+\infty} \mathrm{d}m~n(m)~b(m)~my_\star(k|m)\right]\left[ \int_0^{+\infty} \mathrm{d}m~n(m)~b(m)~my_\mathrm{g}(k|m)\right],
\end{equation}
respectively.

\subsection{Total matter power spectrum}

\begin{figure}
	\centering
	\includegraphics[width=0.70\hsize]{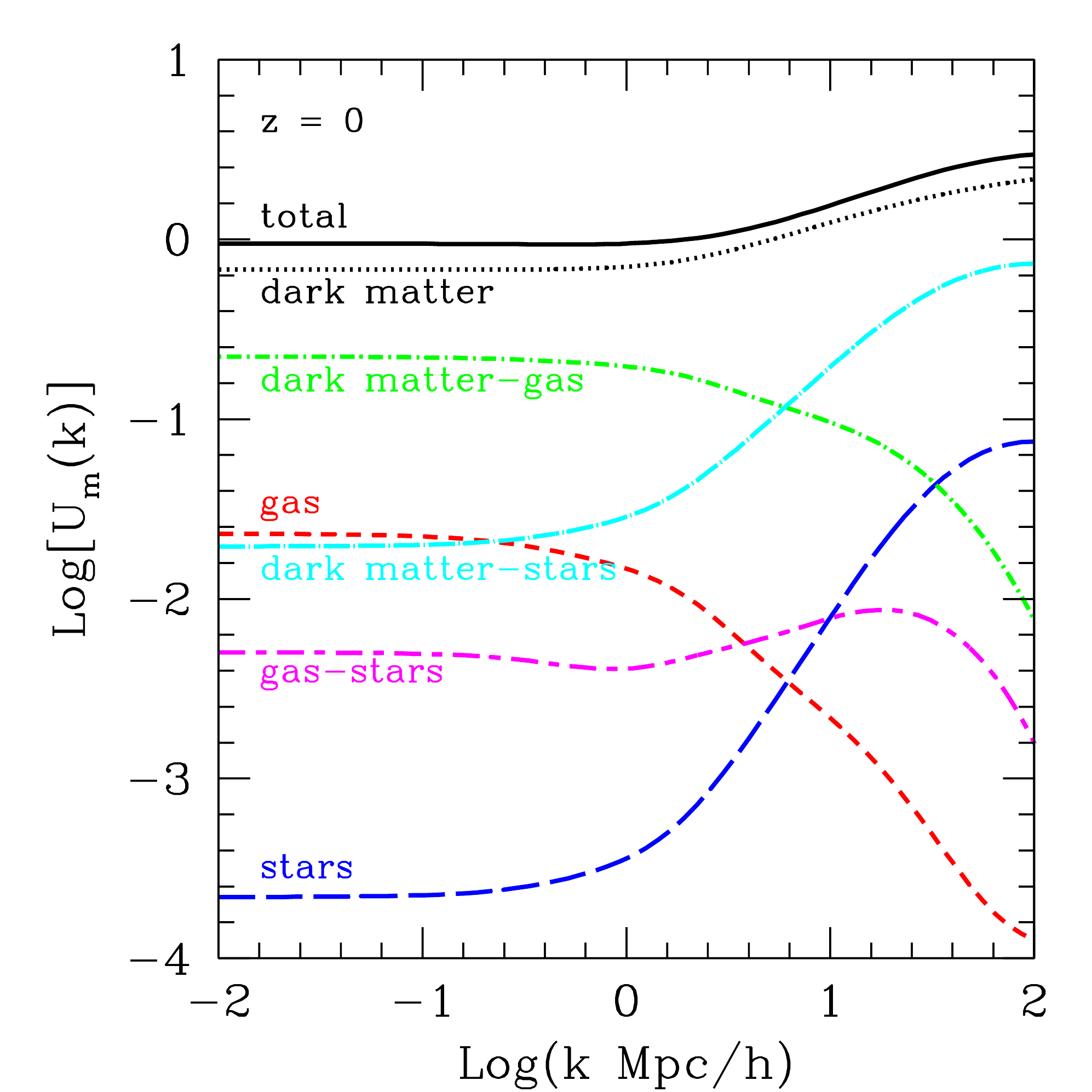}
	\caption{The total matter power spectrum, decomposed in its six separate contributions, as labeled. Each contribution is multiplied by the appropriate prefactor shown in Eq. (\ref{eqn:matterPower}), so that the total power spectrum (black solid line) is just the sum of the contributions shown.}
	\label{fig:matterPower}
\end{figure}

By working out the clustering statistics it is now easy to see that the total matter power spectrum is given by a simple combination of the power and cross spectra described above. Specifically, it can be written as

\begin{equation}\label{eqn:matterPower}
P_\mathrm{m}(k) = \frac{1}{\bar\rho_\mathrm{m}^2}\left[\bar\rho_\mathrm{DM}^2 P_\mathrm{DM}(k) + \bar\rho_\mathrm{g}^2 P_\mathrm{g}(k) + \bar\rho_\star^2 P_\star(k) + 2\bar\rho_\mathrm{DM}\bar\rho_\mathrm{g} P_{\mathrm{DM}\mathrm{g}}(k) + 2\bar\rho_\mathrm{DM}\bar\rho_\star P_{\mathrm{DM}\star}(k) + 2\bar\rho_\star\bar\rho_\mathrm{g} P_{\star\mathrm{g}}(k)\right]~.
\end{equation}
 It is worth noting that in the baryon-free case $\bar\rho_\mathrm{g} = \bar\rho_\star = 0$, and $\bar\rho_\mathrm{DM} = \bar\rho_\mathrm{m}$. It follows that $P_\mathrm{m}(k) = P_\mathrm{DM}(k)$, as it should be.

In Figure \ref{fig:matterPower} I show the total matter power spectrum together with its six separate contributions. Each contribution is correctly weighted according to the respective mean densities, as in Eq. (\ref{eqn:matterPower}). Overall, the total matter power spectrum mirrors the behavior of the DM spectrum, with other matter components producing relatively small corrections. In Figure \ref{fig:matterPower} one can appreciate the relative contribution of different baryonic components as a function of scale. On large scales, $k\lesssim 0.2 h$ Mpc$^{-1}$ the gas contributes for $\sim 3\%$ of the total matter clustering, while stars give a negligible contribution. The cross-correlation between DM and gas is quite substantial ($\sim 20\%$) down to $k\sim 1h$ Mpc$^{-1}$, meaning that the bias $b_\mathrm{d}$ of the diffuse component can be inferred by looking at those scales. At large wavenumbers the stellar clustering becomes the dominant baryonic contribution. The stars contribute $\sim 3\%$ of the total matter power spectrum at $k\gtrsim 30 h$ Mpc$^{-1}$, while on the same scales the impact of the cross-correlation between DM and stars is at the level of $\sim 30\%$. These numbers are obviously going to change under changes in the background cosmology and in the implementation of hydrodynamics, however the general qualitative behavior is the same observed in numerical cosmological simulations.

\begin{figure}
	\centering
	\includegraphics[width=0.70\hsize]{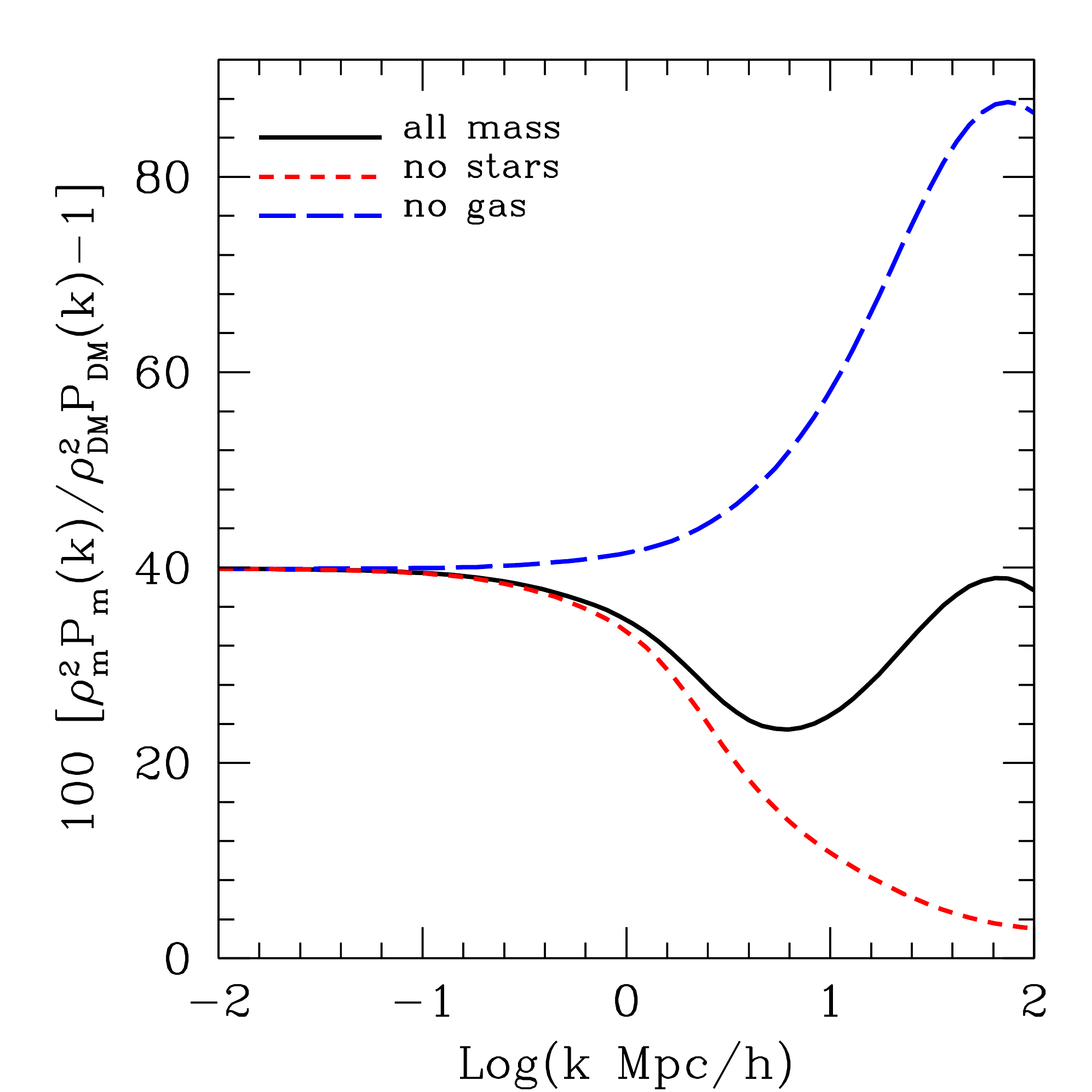}
	\caption{The fractional difference between the total mass power spectrum and the power spectrum of DM, the latter being weighted by the relative DM abundance (black solid line). The red short-dashed line shows the result obtained by ignoring the contribution of stars, where the removed stars are replaced by an equal mass in DM and gas, proportionally to their respective abundances. The blue long-dashed line shows the result obtained by ignoring the contributions of hot gas instead, with a similar replacement.}
	\label{fig:matterPower_PERCENTAGE}
\end{figure}

In order to better highlight the contribution of baryons to the matter clustering, in Figure \ref{fig:matterPower_PERCENTAGE} the DM power spectrum (multiplied by the appropriate prefactor, see Eq. \ref{eqn:matterPower}) has been subtracted off the total mass power spectrum. Also shown, are the results obtained by ignoring the contributions of hot gas and stars, represented by a blue and red curve, respectively. I renormalized the latter two curves to coincide with the black line at large scales. This means assuming that the removed mass (either in gas or stars) is always replaced by an equal mass in the two remaining components (DM and stars or DM and gas), proportionally to their respective abundances. As can be seen there is a distinct effect of baryons on the shape of the overall matter clustering. Specifically, the presence of gas tends to suppress the matter power at intermediate scales, $k\gtrsim 1 h$ Mpc$^{-1}$. This effect is partly counteracted by the presence of stars, which enhance the clustering again at small scales, $k\gtrsim 10 h$ Mpc$^{-1}$, resulting in an overall suppression of $\sim 15\%$ at $k\sim 5 h$ Mpc$^{-1}$. This change in shape obviously depend on the assumed abundance and distribution of baryons.

\section{Theoretical uncertainties of the model}\label{sct:uncertainties}

The SAM developed in this paper rests on four pillars: $i)$ the mass fractions of the different matter components within bound structures; $ii)$ the respective density profiles (as a function of mass and redshift); $iii)$ the baryon-free DM halo mass function; $iv)$ the baryon-free DM linear halo bias. While our knowledge of the first two elements relies mainly on detailed observations of large samples of galaxies and galaxy clusters, this is not the case for the latter two. Indeed, since the real Universe \emph{is} filled with baryons, the DM-only mass function and bias can be known only through $n-$body numerical simulations. In Paper II it is shown how the mass fractions and density profiles change in simulations implementing different kind of baryonic physics, and how this change translates into a change of the different matter power spectra. In this Section instead I discuss the uncertainties in the matter power spectra stemming from our imperfect theoretical knowledge of the halo mass function and linear bias.

\begin{figure}
	\centering
	\includegraphics[width=0.49\hsize]{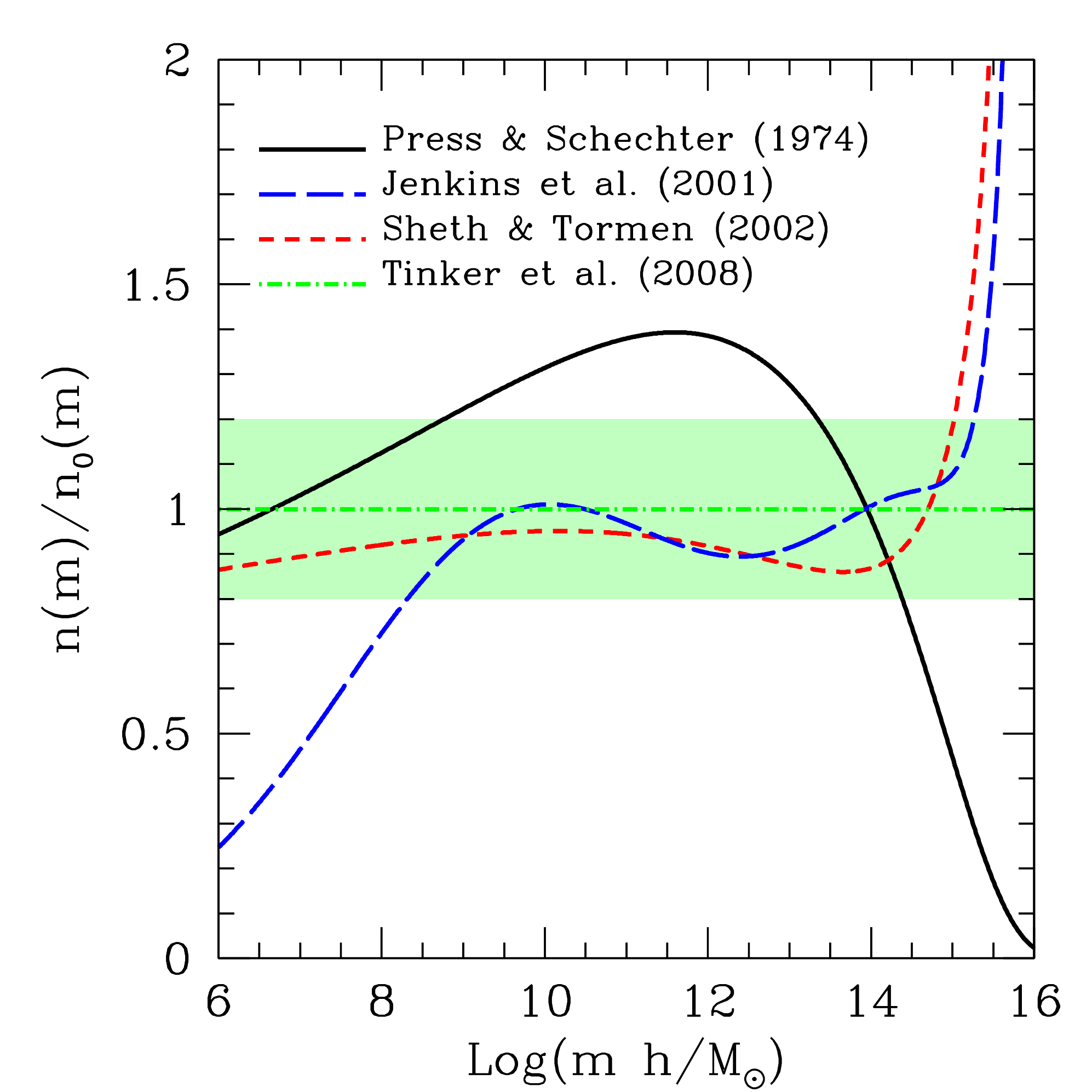}
	\includegraphics[width=0.49\hsize]{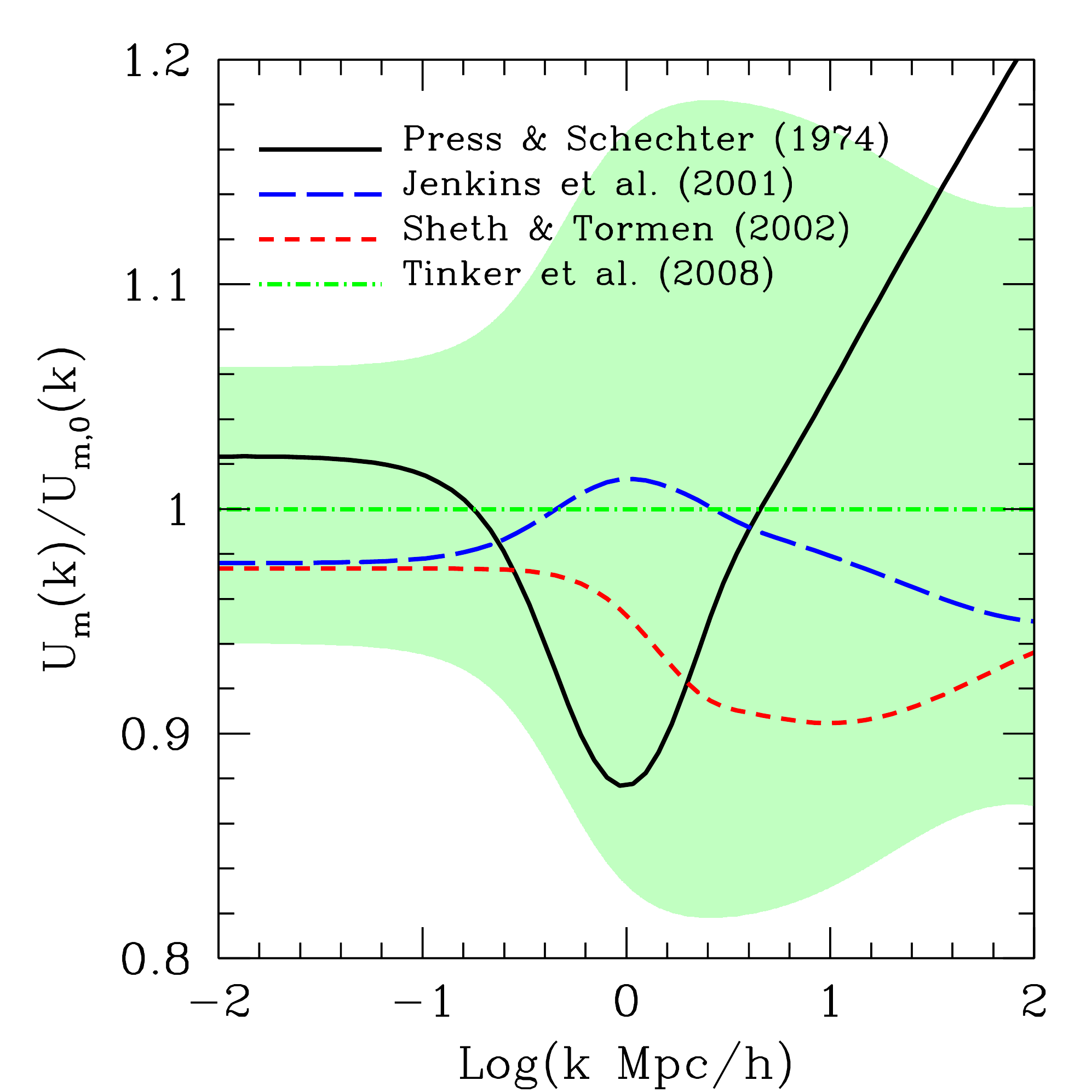}
	\caption{\emph{Left panel}. The DM halo mass function computed according to different prescriptions, as labeled, referred to the Tinker et al. (2008) formula \cite{TI08.1}, which is considered to be the reference one. The green band represents a constant $\pm 20\%$ uncertainty in the reference mass function. \emph{Right panel}. The total matter power spectrum computed with the SAM model by adopting the mass function prescriptions shown in the left panel. All is considered at $z=0$.}
	\label{fig:matterPower_MF}
\end{figure}

Recent works on numerical $n-$body simulations show that the uncertainty on these two theoretical ingredients range from $\sim 10-15\%$ for galaxy-sized halos up to $20-25\%$ for cluster-sized halos \cite{TI08.1,TI10.1,MU13.1}. I found that the various matter power spectra are more affected by uncertainties on the mass function rather than the bias. In part this is due to the fact that the linear halo bias enters only in the $2-$halo parts, and hence its effect tends to die out at intermediate scales. For this reason I focused here only on the uncertainties on the mass function, considering them to be the dominant theoretical sources of error. In the left panel of Figure \ref{fig:matterPower_MF} I illustrate the ratios of mass functions computed according to various prescriptions to the reference mass function $n_0(m)$, that here I assumed to be the Tinker et al. prescription \cite{TI08.1}. Other prescriptions include the classic Press \& Schechter analytic recipe \cite{PR74.1}, its ellipsoidal collapse-based improvement according to Sheth \& Tormen \cite{SH99.1,SH01.1,SH02.1}, and the fit to numerical simulation presented by Jenkins et al. \cite{JE01.1}. The green band represents a constant $\pm 20\%$ uncertainty on the reference mass function. The abundance of DM halos varies widely depending on the prescription adopted. Both the Jenkins et al. and the Sheth \& Tormen formulae are roughly in agreement with the reference mass function at intermediate masses, however they deviate substantially at very low and very high masses. The Press \& Schechter prescription deviates substantially from the reference mass function at al masses. However, this latter prescription is considered outdated and too simplistic, thus it is shown only for the purpose of illustration and not discussed further.

\begin{figure}
	\centering
	\includegraphics[width=0.32\hsize]{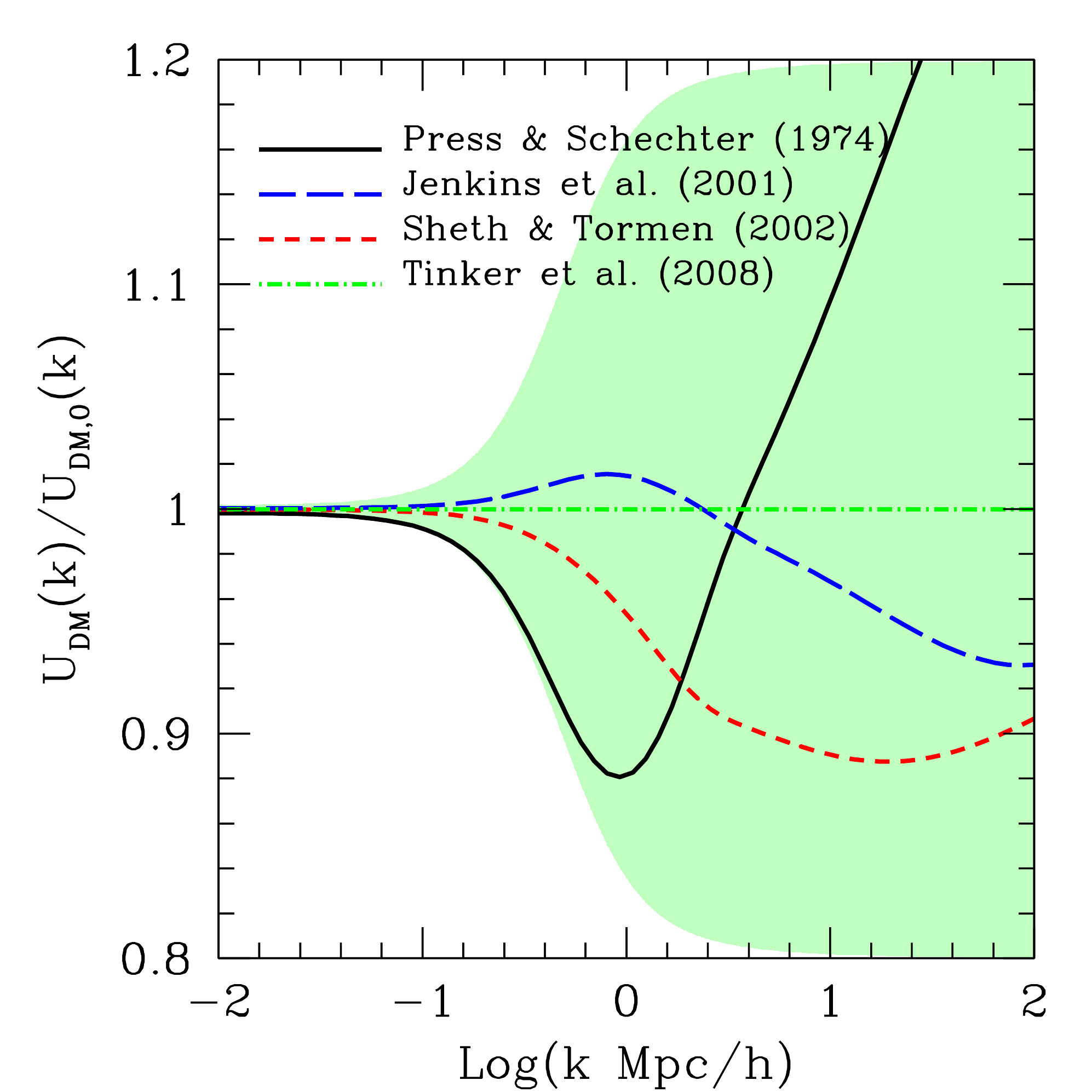}
	\includegraphics[width=0.32\hsize]{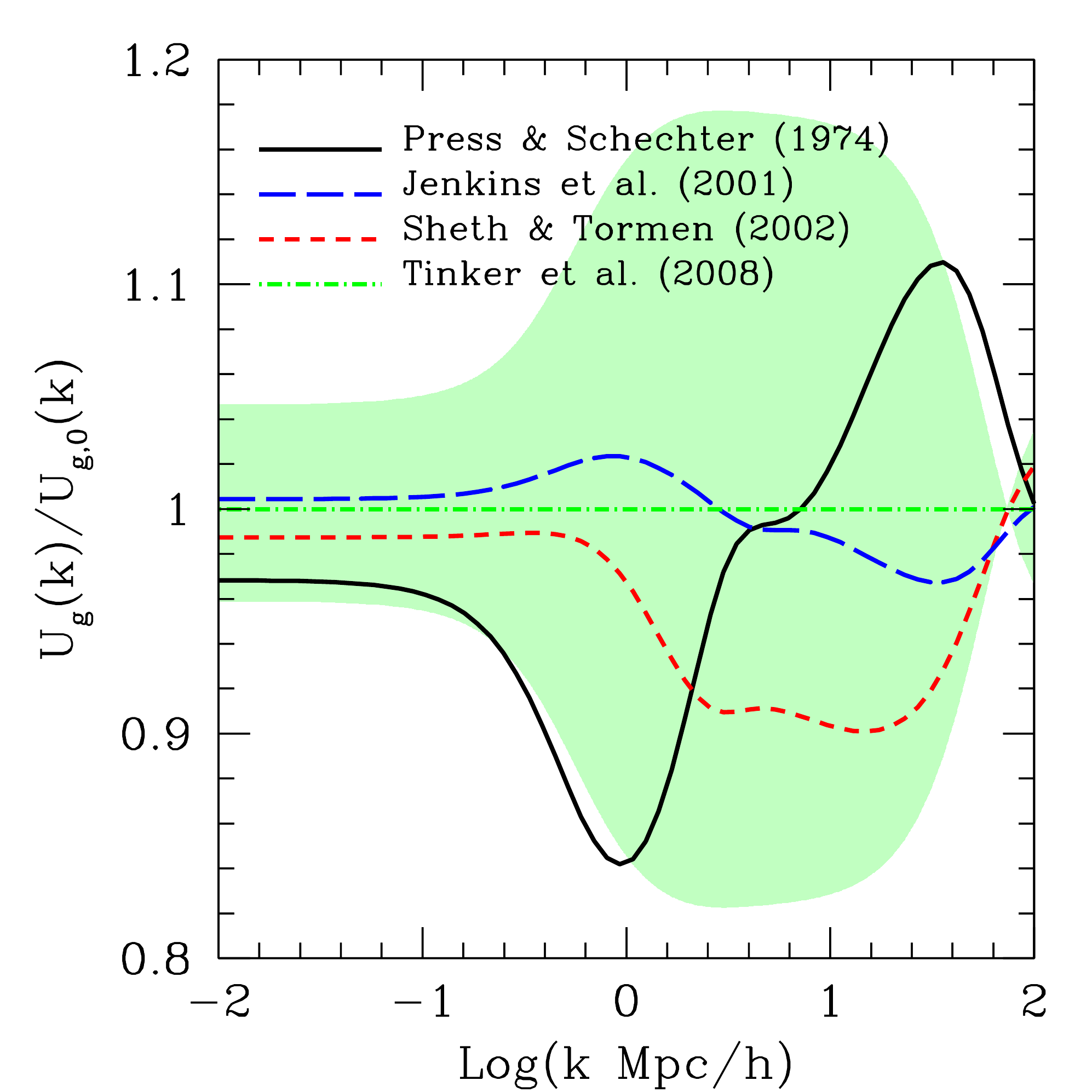}
	\includegraphics[width=0.32\hsize]{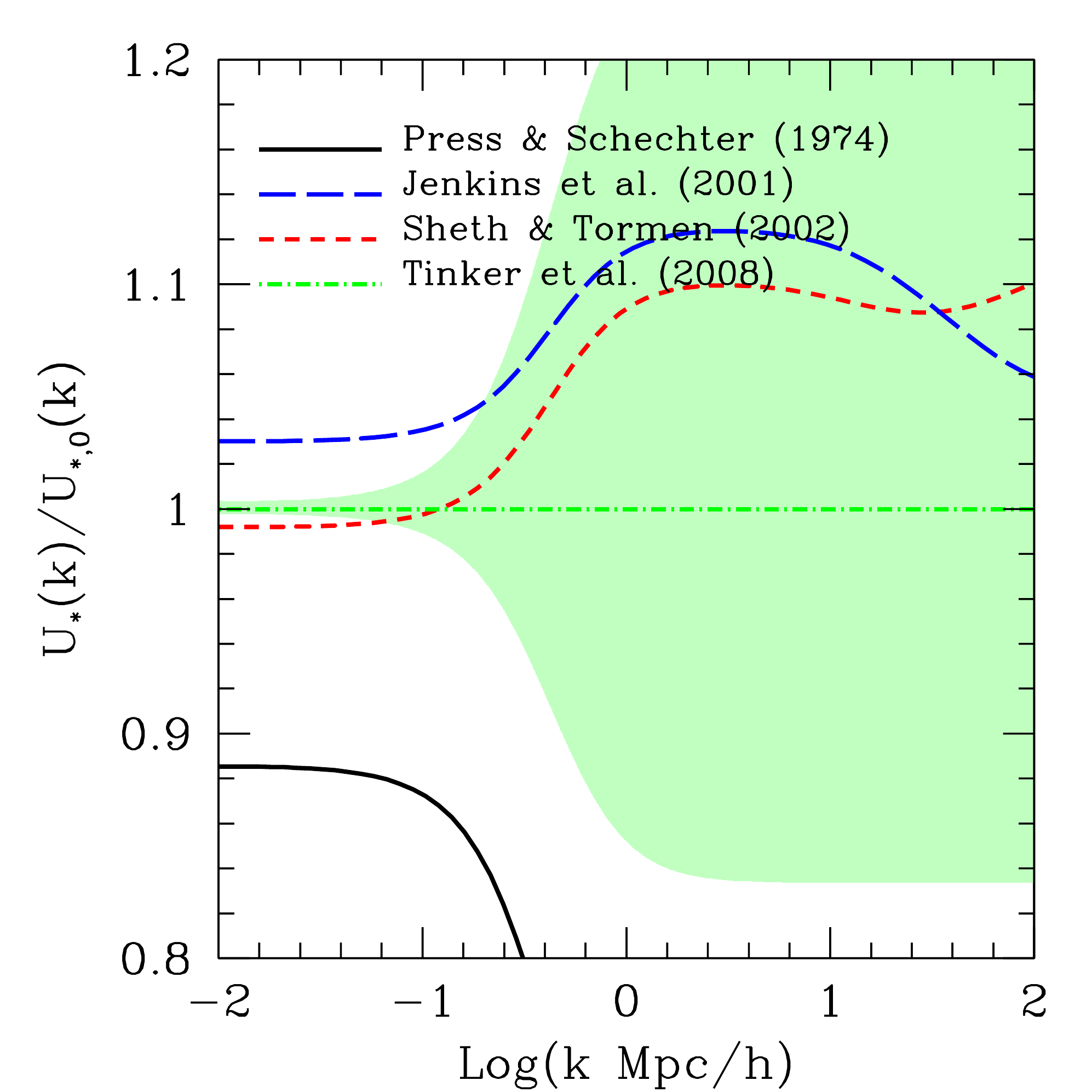}
	\caption{The same as the right panel of Figure \ref{fig:matterPower_MF} but referring only to the DM power spectrum (left panel), the gas power spectrum (middle panel), and the stellar power spectrum (right panel).}
	\label{fig:darkMatterPower_MF}
\end{figure}

The right panel of Figure \ref{fig:matterPower_MF} shows the total matter power spectrum computed adopting the different mass function prescriptions illustrated in the left panel. As can be seen, the effect of the uncertainty in the mass function prescription is substantial. By switching from the reference formula to other prescriptions based on numerical simulations, the total matter power spectrum changes by $\sim 5-10\%$. By assuming a constant $\pm 20\%$ uncertainty in the reference mass function one obtains deviations up to $\sim 20\%$ in the matter power. These deviations are of the same order of magnitude of the effect of baryons that one is trying to study. Similar large impact of uncertainties in the halo mass function on semi-analytic models of the LSS have recently been found by Valageas \cite{VA13.1}. In Section \ref{sct:discussion} below we elaborate on the consequences of these uncertainties. In Figure \ref{fig:darkMatterPower_MF} I show the impact of mass function uncertainties on the power spectra of each individual matter component. The same general conclusions apply, however it is worth noticing that at large scale there is no uncertainty on the DM power spectrum or on the stellar power spectrum (the latter only when the shift in the mass function is constant). The former is due to the fact that the DM spectrum is always renormalized in a way such that it tends to the linear matter power spectrum on large scales (see Eq. \ref{eqn:constraint}). The latter is due to the fact that whenever the mass function is shifted by a constant factor, the normalization of the stellar fraction gets adjusted by the same shift in order to leave the mean stellar density unchanged. The uncertainty on the large-scale gas power spectrum is also particularly important because it propagates into an uncertainty in the determination of the bias of the diffuse component. I verified that changing the mass function by a constant $20\%$ implies an error in the determined $b_\mathrm{d}$ of $\sim 3\%$.

\section{Discussion and conclusions}\label{sct:discussion}

This paper presents a generalized SAM that allows one to describe the clustering of DM and baryons. The SAM is based on the halo model (\cite{CO02.2} and references therein) and as such it evaluates the power spectrum of a given mass component by convolving the linear matter clustering with the average halo density run of that component, subsequently averaging over the halo mass function and linear bias. This SAM has been extensively tested in the literature for DM clustering, and here it has been straightforwardly  extended to the stellar clustering. The clustering of gas density fluctuations has also been addressed here, however it requires some additional complications. Because the vast majority of the hot gas in the Universe is placed outside gravitationally bound structures, the halo model requires the introduction of two additional contributions to the gaseous power spectrum, related with the large-scale diffuse component. Eventually, the total mass power spectrum can be estimated by combining together the spectra of each component and their mutual cross spectra. The total mass power spectrum is what it is measured by, e.g., cosmic shear.

In order to illustrate the method I computed the clustering of stars and hot gas, by adopting educated guesses for the model parameters. The following results can be emphasized.

\begin{itemize}
\item The power spectrum of gas at large scales $k\lesssim 1h$ Mpc$^{-1}$ is dominated by density fluctuations in the diffuse component. The contribution due to the clustering of individual structures is of $\sim 10\%$ only.
\item On intermediate/small scales the dominant contribution to the gaseous power spectrum is given by density fluctuations within individual objects, and hence it is sensitive to the gas distribution inside host DM halos.
\item The stellar power spectrum is dominated by the mutual clustering of structures for $k \lesssim 0.4h$ Mpc$^{-1}$, while it becomes driven mainly by the stellar distribution within separate structures on smaller scales.
\item Besides DM, the cross-correlation between DM and gas and that between DM and stars provide the highest contributions to the total mass clustering. The former is prominent at large and intermediate scales ($k \lesssim 6 h$ Mpc$^{-1}$), while the latter is at smaller scales.
\item It follows that, for a given DM distribution, the total mass clustering is expected to be sensitive to the bias of the diffuse gas component at large scales, and to the distribution of stars within DM halos at small scales.
\item The largest source of theoretical uncertainty in the SAM is given by our imperfect knowledge of the DM halo mass function. By changing the mass function prescription the total mass power spectrum can oscillate by up to $\sim 10\%$.
\end{itemize}
While the impacts of mass function uncertainties on semi-analytic modeling of structure formation are seldomly addressed (see however \cite{VA13.1}), they are bound to affect any SAM. Therefore, such uncertainties should be always kept in mind when making predictions based on these models. In the present case, our imperfect knowledge of the DM halo mass function alters the total mass clustering by an amount that is a substantial fraction of the effect of baryonic matter. One way to take this into account, which will be implemented in the future, is to parametrize this uncertainty and marginalize over it when estimating cosmology or the parameters of baryonic physics.

Following this reasoning, the implementation of the SAM presented here is by no means final. Rather, there are a number of ways in which the model can be improved, and whether such improvements are necessary or not in view of future cosmic shear surveys can only be assessed by extensive testing of the method, a beginning of which will be presented in Paper II. Apart from the marginalization over mass function uncertainties mentioned above, some other possible refinements are as follows.

\begin{itemize}
\item The halo model is known to lack some accuracy at the transition between the $1-$halo and the $2-$halo terms \cite{RE02.2,SE03.2}, which can be mitigated by the heuristic halo exclusion or some ad-hoc scale-dependence of the linear bias \cite{TI05.1,VA13.3}. This will be discussed in Paper II.
\item The assumption that the bias of the diffuse gas component is a constant should be a good approximation on very large scales, but it probably breaks down at some intermediate scale, where the diffuse component is still relevant. This will also be discussed in Paper II, and some improvement to the modeling of this component could be devised in the future.
\item The matter density profiles have been assumed to be universal, with only their parameters changing with structure mass. This is probably not an extremely good approximation for gas and stars, so that some more sophisticated modeling of the mass distributions within DM halos should be adopted in future studies.
\item The assumption that profile parameters are simple power-laws of the structure mass might not be accurate enough, especially in light of the fact that the halo model needs integrating over a large range of masses. This could be adjusted by assuming some kind of non-linear dependency, however it would come to the cost of increasing the number of model parameters.
\item The stochasticity of mass density profiles could be taken into account by convolving the power spectra resulting from the SAM with probability distributions for the model parameters. This procedure would also increase the number of parameters, so it has been decided to not include it yet.
\end{itemize}

On a related note, the number of free parameters that the SAM developed in this paper uses to describe the mass distribution within bound structures amounts to $10$. By including also the bias of the diffuse gas component and the way in which the gas and stellar fractions change with mass, this raises up to $15$ parameters. While this large number of parameters is unavoidable if one wants to have an internally consistent description of the total matter power spectrum, this can lead to difficulties in the inverse application of the model that has been mentioned in the Introduction, namely exploiting high-precision cosmic shear surveys in order to infer the behavior and physics of baryons. One possibility to reduce the dimensionality of the problem could be to held fixed the parameters that are better known (e.g., the DM halo concentration) and leave only the most uncertain ones as free to vary. However, a viable and more elegant solution would be to use Monte-Carlo techniques in the context of Bayesian inference in order to derive posterior probability distributions for all the model parameters, as well as their covariances. Recently Lu and collaborators \cite{LU11.1,LU12.1,LU13.2} adopted a similar approach in order to infer galaxy evolution model parameters based on observed luminosity functions, obtaining interesting constraints on $13-18$ parameters of the model.

To conclude, this paper describes a SAM based on the halo model that will be useful for the interpretation of future cosmic shear observations. Besides aiding in understanding the behavior of baryonic components that are difficult to access through other observations, such as the WHIM and the intra-halo stars, this method will also be useful for cosmological parameter estimation and for better understanding the contribution of galaxy formation physics on the assembly of the LSS.

\section*{Acknowledgments}

I wish to thank L. Moscardini and H. Hoekstra for useful insights on this manuscript. The research leading to these results has received funding from the European Commission Seventh Framework Programme (FP7/2007-2013) under grant agreement n$^\circ$ 267251. I acknowledge financial contribution from the University of Florida through the Theoretical Astrophysics Fellowship. I am also grateful to an anonymous referee for useful comments that improved the presentation of this work.

{\small
\bibliographystyle{JHEP}
\bibliography{./master}
}

\end{document}